\documentclass{svjour2}
\journalname{}

\usepackage{bm}
\usepackage{cite}
\usepackage{amsmath}
\usepackage{amsfonts}
\usepackage{graphicx}

\numberwithin{equation}{section}

\newcommand{\flow}[1]{\phi^{#1}}
\newcommand{\tanflow}[2]{D_{#1}\phi^{#2}} 
\newcommand{\dxi}{\delta\xi}
\newcommand{\deta}{\delta\eta}
\newcommand{\lam}[1]{\lambda^{(#1)}}
\newcommand{\Vf}[2]{V^{(#1)}_{#2,+}}
\newcommand{\Uf}[2]{U^{(#1)}_{#2,+}}
\newcommand{\Vr}[2]{V^{(#1)}_{#2,-}}
\newcommand{\Ur}[2]{U^{(#1)}_{#2,-}}
\newcommand{\W}[2]{W^{(#1)}_{#2}}
\newcommand{\A}{\mathcal{A}}
\newcommand{\N}{\mathcal{N}}
\newcommand{\D}{\mathcal{D}}
\newcommand{\K}{k_x}
\newcommand{\BXiJ}{\Big(\Xi \Big|J_{4N}\Xi\Big)} \newcommand{\XiJ}{(\Xi |J_{4N}\Xi)}
\newcommand{\C}{\mbox{c}}
\newcommand{\s}{\mbox{s}}
\newcommand{\Nd}{N_{\Omega}^t}

\begin{document}

\title {Lyapunov Mode Dynamics in Hard-Disk Systems} 

\author{D J Robinson \and G P Morriss}
\institute{School of Physics, University of New South Wales, 
Sydney 2052, Australia \email{djr233@cornell.edu, G.Morriss@unsw.edu.au}} \date{\today}

\maketitle

\begin{abstract}
The tangent dynamics of the Lyapunov modes and their dynamics as generated numerically - {\it the numerical dynamics} - is considered. We present a new phenomenological description of the numerical dynamical structure that accurately reproduces the experimental data for the quasi-one-dimensional hard-disk system, and shows that the Lyapunov mode numerical dynamics is linear and separate from the rest of the tangent space. Moreover, we propose a new, detailed structure for the Lyapunov mode tangent dynamics, which implies that the Lyapunov modes have well-defined (in)stability in either direction of time. We test this tangent dynamics and its derivative properties numerically with partial success. The phenomenological description involves a time-modal linear combination of all other Lyapunov modes on the same polarization branch and our proposed Lyapunov mode tangent dynamics is based upon the form of the tangent dynamics for the zero modes. \end{abstract}

\section{Introduction}

\bibliographystyle{spmpsci}

We consider the dynamics of the Lyapunov modes found in two-dimensional hard-disk systems. These modes are a subset of the so-called Lyapunov vectors, which are particular phase space perturbations whose (in)stability is characterized by a Lyapunov exponent. The modes themselves exhibit spatial and time modality and study of the Lyapunov modes has been consequently mainly motivated by their possible connection to hydrodynamic fluctuations in fluids. 

The empirical observation of these modes \cite{DPH96,MPH98,PH00} has led to various studies of their physical origins based on several different approaches. Firstly, random matrix approximations have demonstrated the importance of translational invariance in the phase space \cite{EG00}, while a combined random matrix and master equation approach \cite{TM02} has predicted the existence of the Lyapunov steps, which are the sets of degenerate Lyapunov exponents characterizing the Lyapunov mode (in)stability, and of the spatial modality. Alternatively, application of kinetic theory in Ref. \cite{MM01} exhibited a connection between mode generation and conservation laws, and predicted polarization of the modes into longitudinal and transverse polarization branches. Further, describing Lyapunov modes as Goldstone modes \cite{WB04} arising from symmetry breaking produced approximations to the wavenumber dependence of the Lyapunov exponents and the form of the modes. 

The abovementioned approaches focus mainly on the generation and spatial structure of the Lyapunov modes. The underlying general theory of the tangent dynamics of the Lyapunov modes is well-known (see e.g. \cite{EFPZ05} or \cite{ER85} for a review), while study of the dynamics of the Lyapunov modes as they are generated numerically - the numerical dynamics - has included: observation of time-periodicity in the longitudinal modes \cite{PH00,MP02,FHPH04,TM05}; prediction of the consequent mode propagation speeds for low densities (see e.g. Refs. \cite{MM01,WB04}); and has revealed a connection between the time oscillating period and the momentum autocorrelation function \cite{TM05}. Refs. \cite{EFPZ05,TM05} have considered the numerical dynamics of longitudinal modes in terms of rotations in the tangent space, but the numerical dynamical structure of the modes has not been examined in the same level of detail as the spatial structure or the degeneracy of the Lyapunov exponents. 

In the present work, we present a phenomenological description of the numerical dynamical structure of modes in the quasi-one-dimensional system, which accurately reproduces the numerical dynamics of its Lyapunov modes. This description demonstrates that the mode numerical dynamics is linear and appears to separate from the numerical dynamics of the other, non-modal Lyapunov vectors. Moreover, we propose for the first time a detailed structure of the mode tangent dynamics. This mode tangent dynamics possesses a symplecticity property, which is shown to imply that the Lyapunov modes lie in the so-called Oseledec decomposition spaces, and so have well-defined (in)stability in either direction of time. We test this tangent dynamics and its derivative properties numerically with partial success. The essential feature of the phenomenological description is that the numerical dynamics of a Lyapunov mode is a time-modal linear combination of all other modes on the same polarization branch. Our mode tangent dynamics is based on this phenomenological description and an extension of the zero mode tangent dynamics. 

\section{Lyapunov Vectors and Modes}

\subsection{Tangent Space Dynamics}
\label{sec:GPTSD}
Consider an $r$-dimensional system of $N$ identical particles, with phase space $M \simeq \mathbb{R}^{2rN}$. For a state $\xi \in M$ the dynamics $\xi(t)$ is represented by the phase flow $\flow{t} : M \to M$. That is, $\xi(t) = \flow{t}(\xi)$, where \begin{equation}
	\label{eqn:PSD}
	\xi = (p,q) = (p_1,\ldots,p_N,q_1,\ldots,q_N) \end{equation}
and $p_i$ ($q_i)$ is the momentum (position) of the $i$th particle. 

In the present work we consider a system of hard-spheres, which interact only by instantaneous elastic collisions. There is no external potential in this system and we neglect tangent collisions so that this phase flow is differentiable. A small perturbation $\dxi$ of the state $\xi$ then has dynamics (to first order) \begin{equation}
\label{eqn:TFD}
\dxi(t) = \tanflow{\xi}{t}\dxi,
\end{equation}
where $\tanflow{\xi}{t} \equiv \partial \flow{t}/\partial \xi$ is the $2rN \times 2rN$ matrix of partial derivatives of the phase flow. The map $\tanflow{\xi}{t}$ is called the tangent flow, and we may consider $\tanflow{\xi}{t} : T_{\xi}M \to T_{\flow{t}(\xi)}M$, where $T_{\xi}M$ is the tangent space to $M$ at $\xi$. The dynamics in Eq. ({\ref{eqn:TFD}) therefore implies a perturbation $\dxi$ may be considered to be a tangent vector, which is an element of the tangent space. We henceforth adopt the convention of calling $\dxi$ a tangent vector. The components of $\dxi$ are written \begin{equation}
\label{eqn:DTVC}
\dxi = (\delta p,\delta q) = \Big(\delta p_1,\ldots,\delta p_N, \delta q_1, \ldots, \delta q_N\Big)~. \end{equation}

\subsection{Lyapunov Exponents and Vectors} \label{sec:LMLEV}
A detailed summary of the following theory of Lyapunov vectors and the numerical scheme is presented in Appendix \ref{sec:TF}. Firstly, it is well-known that one may choose an orthogonal basis of the tangent space such that in the positive time limit $t \to \infty$, each basis element $\dxi^+_k$ satisfies \begin{equation}
	\label{eqn:DLE}
	\|\dxi^+_k(t)\| \to \exp\big(\lam{j_k}t\big)\|\dxi^+_k\|~, \end{equation}
for $k=1,\ldots,2rN$ and $1\le j_k \le l$ (see App. \ref{sec:TFTSD}). Here $l \le 2rN$ is the number of distinct exponents $\lam{j_k}$, which are called Lyapunov exponents and are ordered such that $\lam{1} >\ldots>\lam{l}$. The ordered set of Lyapunov exponents is called the Lyapunov spectrum, and we say the tangent vector $\dxi^+_k$ is `associated with' the Lyapunov exponent $\lam{j_k}$. 

The notation $\lam{j}$ here denotes a distinct Lyapunov exponent. A particular exponent $\lam{j}$ may have multiplicity $m(j)$, such that $j_k = \ldots = j_{k+m(j)} \equiv j$. The associated basis elements $\dxi^+_k,\ldots,\dxi^+_{k+m(j)}$ then span a subspace of $T_{\xi}M$, whose elements are themselves all associated with $\lam{j}$. There are therefore many possible (pairwise linearly independent) choices of $\dxi^+_k,\ldots,\dxi^+_{k+m(j)}$ for $m(j) \ge 2$.We label this subspace $\Uf{j}{\xi} \subset T_{\xi}M$, and call the elements of $\Uf{j}{\xi}$ Lyapunov vectors. 

The system under consideration is Hamiltonian, so that the dynamics $\flow{t}$ is reversible and we may similarly consider the behavior of tangent vectors in the negative time limit, $t \to -\infty$ (see Sec. \ref{sec:TFHP}). In this case, the Lyapunov spectrum is $-\lam{l}>\ldots-\lam{1}$, and we may choose another orthogonal basis of $T_{\xi}M$ such that \begin{equation}
	\label{eqn:DLE2}
	\|\dxi^-_k(t)\| \to \exp\big(-\lam{j_k}t\big)\|\dxi^-_k\|~. \end{equation}
There similarly exist subspaces $\Ur{j}{\xi} \subset T_{\xi}M$ whose elements are associated to $-\lam{j}$ for $t \to -\infty$. 

Notably, a Lyapunov vector associated to $-\lam{j}$ for $t \to -\infty$ is not necessarily associated to $\lam{j}$ for $t \to +\infty$, so that $\Uf{j}{\xi} \not= \Ur{j}{\xi}$ in general. However, it is possible to decompose the tangent space into subspaces whose elements do have well-defined Lyapunov exponents in either direction of time (see App. \ref{sec:TFCS}). These subspaces are called Oseledec spaces, denoted $\W{j}{\xi}$, and a tangent vector $\dxi \in \W{j}{\xi}$ satisfies both Eqs. (\ref{eqn:DLE}) and (\ref{eqn:DLE2}) with exponent $\lam{j}$. The Oseledec spaces are also covariant: they are preserved by the tangent dynamics, so that $\tanflow{\xi}{t}\W{j}{\xi} = \W{j}{\flow{t}(\xi)}$. 

\subsection{Numerical Scheme}
\label{sec:LVMNS}

A set of $2rN$ orthonormal Lyapunov vectors may be generated numerically by the numerical scheme of Benettin \cite{BGGS80} and Shimada \cite{SN79}, which is used primarily to generate the Lyapunov spectrum. This scheme (see App. \ref{sec:TFNS}) involves evolving a set of randomly chosen, linearly independent tangent vectors forwards in time under the tangent flow, whilst periodically performing Gram-Schmidt orthonormalization. The resulting set converges to an orthonormal Lyapunov vector basis. Importantly, the Lyapunov vectors generated by this scheme are strictly associated with $-\lam{j}$ for $t \to -\infty$, rather than with $+\lam{j}$ for $t \to +\infty$ as would be expected naively. 

The Hamiltonian nature of the system further guarantees (see App. \ref{sec:TFFHP}) that there is always a Lyapunov exponent equal to zero (that has even multiplicity) and all other distinct exponents are paired signwise, so \begin{equation}
	\label{eqn:LEPS}
	\lam{j} = -\lam{l-j+1}.
\end{equation}
It follows that the number of distinct exponents $l$ is always odd and that $\lam{[l+1]/2} = 0$. 

Furthermore, a basis of $2rN$ Lyapunov vectors may be uniquely specified by considering only the first $rN$ Lyapunov vectors generated by the numerical scheme (see App. \ref{sec:TFFHP}). These $rN$ Lyapunov vectors span $U_{\pm,\xi}^{(1)}\oplus\cdots\oplus U_{\pm,\xi}^{([l-1]/2)}$ together with half of $U_{\pm,\xi}^{([l+1]/2)}$. We henceforth call these Lyapunov vectors the positive Lyapunov vectors, since they are associated with exponents $\lambda \ge 0$. The remaining $rN$ Lyapunov vectors associated with $\lambda \le 0$ are called negative Lyapunov vectors. 

\subsection{Quasi-One-Dimensional System} 

Henceforth we consider only two-dimensional rectangular systems of $N$ particles with rectangular boundaries, so in Eq. (\ref{eqn:PSD}) the components $p_i,q_i \in \mathbb{R}^2$. We use Cartesian coordinates $(x,y)$ in this system and let $q = (x,y)$ and $p = (p_x,p_y)$. For simplicity, we will mainly consider a special case of the rectangular system called the quasi-one-dimensional system. This system is a two-dimensional system which is sufficiently narrow so that particles cannot pass one another (see Fig. \ref{fig:QDS}). 

\begin{figure}[ht]
\begin{center}
	\includegraphics{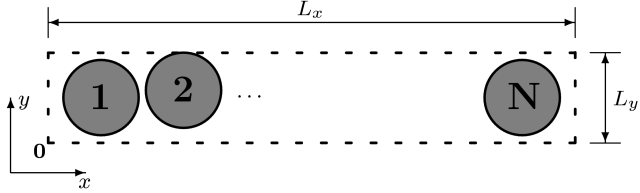}
\end{center}
	\caption{Schematic of an $N$ hard-disk quasi-one-dimensional system. The height $L_y$ is sufficiently small that the disks cannot pass one another. We choose the coordinate origin to be located at the bottom left corner of the system, and system boundaries are denoted by dashed lines. The pairs of opposite boundaries $x =0,L_x$ or $y=0,L_y$ may possess either periodic (P) or hard (H) boundary conditions. We call the system with hard $x=0,L_x$ boundaries and periodic $y=0,L_y$ boundaries the HP quasi-one-dimensional system.} 	\label{fig:QDS}
\end{figure}

\subsection{Zero Modes}
\label{sec:LMZM}

In the case of periodic boundary conditions, the rectangular system has full translational spatial and time symmetry. There are then (see App. \ref{sec:TFSZM}) six Lyapunov vectors associated with $\lam{[l+1]/2} = 0$ in either time limit. The first of these Lyapunov vectors corresponds to uniform translation in the $x$ direction and is $\dxi_1 = 1/N^{1/2}(0,0,e_1,\ldots,e_1)$, where $e_1 = (1,0)$ is repeated $N$ times and $0$ is the $N$ dimensional zero vector. We simplify this notation by writing (see Eq. \ref{eqn:DTVC}) \begin{align}
	\label{eqn:FCN}
\dxi = \Big(\ldots,(\delta p_{i_x},\delta p_{i_y}),\ldots,\ldots,(\delta q_{i_x},\delta q_{i_y}),\ldots\Big) \equiv (\delta p_x,\delta p_y,\delta q_x,\delta q_y)~, 	 \end{align}
so $\dxi_1 = 1/N^{1/2}(0,0,1,0)$. The other five Lyapunov vectors are then $\dxi_2 = 1/N^{1/2}(0,0,0,1)$, $\dxi_3 = 1/\|p\|(0,0,p_x,p_y)$, $\dxi_4 = 1/N^{1/2}(1,0,0,0)$, $\dxi_5 = 1/N^{1/2}(0,1,0,0)$ and $\dxi_6 = 1/\|p\|(p_x,p_y,0,0)$, which correspond respectively to translation in the $y$ direction, time translation, a uniform momentum boost in the $x$ direction, a uniform momentum boost in the $y$ direction, and an energy shift. Note that in the case of e.g. $\dxi_3$, putting $\delta p_x = p_x$ means that $\delta p_{i_x} = p_{i_x}$. These Lyapunov vectors are collectively called the zero modes. One finds that linear combinations of $\dxi_1$, $\dxi_2$ and $\dxi_3$ ($\dxi_4$, $\dxi_5$ and $\dxi_6$) belong to the positive (negative) Lyapunov vectors (see Sec. \ref{sec:LVMNS}). 

In the case that boundary conditions at one pair of opposite boundaries of the rectangular system are reflecting rather than periodic, then one symmetry is broken and the number of zero modes is reduced by two. For a quasi-one-dimensional system with mixed period and reflecting boundary conditions, we thus expect four zero modes. We assume in the present work that time symmetry is never broken and total energy is always conserved, so there are always at least two zero modes for the system. 

\subsection{Lyapunov Modes in the Quasi-One-Dimensional System} 

Consider a positive Lyapunov vector, $\dxi \in T_{\xi}M$ of unit norm at some fixed time $t$. Generally in any $r$ dimensional system, the Lyapunov vectors generated by the numerical scheme are strongly dynamically localized (see e.g. Ref. \cite{PP98}). This means that for a small, rapidly changing index subset of the integers $1,\ldots,2rN$ the components $\delta p_i, \delta q_i \sim 1$ and $\delta p_i, \delta q_i \ll 1 $ elsewhere. 

However, Lyapunov exponents close to zero exhibit a degenerate step-like structure, and the Lyapunov vectors associated with these exponents are found to be delocalized over the $N$ particles \cite{DPH96,MPH98,PH00}. More precisely, the components $\delta p_i$ ($\delta q_i$) are approximately linear combinations of $\phi_p(p_i)$ ($\phi_q(q_i)$) and $p_i\psi_p(q_i)$ ($p_i\psi_q(q_i)$), where $\phi_{p},\phi_{q}:\mathbb{R}^2 \to \mathbb{R}^2$ and $\psi_{p},\psi_{q}:\mathbb{R}^2 \to \mathbb{R}$ are modal functions. The functions $\phi_{p}$ or $\phi_{q}$ are called modal vector fields and $p_i\psi_p$ or $p_i\psi_q$ are called a scalar modulation of the momentum field. The Lyapunov vectors with this modal property are called Lyapunov modes, and in this work will be denoted by $\deta$. The Lyapunov modes may be represented by the fields $\phi_p$, $\phi_q$, $\psi_p$ and $\psi_q$ which approximate them. We call this the vector field representation of the Lyapunov mode. 

We call the modes belonging to the positive Lyapunov vectors the positive modes. For a general two dimensional system, the zero modes belonging to the positive modes - the positive zero modes - are linear combinations of \begin{align}
	\deta_1 & = \begin{pmatrix}1\\0\end{pmatrix}, & \deta_2 & = \begin{pmatrix}0\\1\end{pmatrix}, & \deta_3 = 	\frac{1}{\|p\|}\begin{pmatrix}p_x\\p_y\end{pmatrix}~.\label{eqn:ZMVFR} \end{align}
Note that we write the positive zero modes in the vector field representation, and we only write the spatial components, as the momentum components are all zero. 

For a general two dimensional system, the modal scalar or vector fields that well-approximate the non-zero Lyapunov modes have a certain number of nodes in the $x$ or $y$ direction. We define $n_x$ ($n_y$) to be the number of nodes in the $x$ ($y$) direction of this Lyapunov mode and define the wavevector of the mode to be \begin{equation}
	\label{eqn:DWV}
k \equiv (k_x,k_y) = \Big(\pi n_x/L_x, \pi n_y/L_y\Big)~, \end{equation}
where $L_x$ and $L_y$ are the spatial sizes of the system. There may clearly be several modes with the same wavenumber $\|k\|$, but for a particular wavenumber, one finds that the corresponding positive modes are associated with Lyapunov exponent \begin{equation}
		\label{eqn:LEDR}
\lambda \approx C_T\|k\|,\ \mbox{or}\ \lambda \approx C_L\|k\|~, \end{equation}
where $C_T$ and $C_L$ are some scalar constants of the system (see e.g \cite{MM01} or \cite{EFPZ05}). The modes associated with $C_L\|k\|$ are called longitudinal modes, while those associated with $C_T\|k\|$ are called transverse modes. Generally there are twice as many longitudinal modes as transverse modes for any particular $\|k\|$: a longitudinal mode may be written as $\alpha\deta_L + \beta\deta_P$, where $\deta_L$ is formed by a modal vector field called an L-mode and $\deta_P$ is a formed by a scalar modulation of the momentum field called a P-mode; while a transverse mode is written as $\deta_T$, a modal vector field called a T-mode. For this reason, the longitudinal (transverse) modes are often referred to as LP-modes (T-modes), and the two linearly independent orthogonal linear combinations of a particular $\deta_L$ and $\deta_P$ are called an LP mode pair. 

One notable empirical feature of the non-zero positive or negative Lyapunov modes is that the momentum component $\delta p$ of a mode is found to be proportional to the spatial component $\delta q$. That is, we may write \begin{equation}
	\label{eqn:SMR}
\deta \approx (C\delta q,\delta q)~,
\end{equation}
where $C$ is a scalar. This relation means that we need only specify the spatial part of the Lyapunov mode in order to describe the entire mode (see e.g. \cite{EFPZ05}), and henceforth we consider only the spatial parts of the mode unless stated otherwise. Equation (\ref{eqn:SMR}) also permits us to visually represent a Lyapunov mode by drawing each (2-dimensional) $\delta q_i$ at the corresponding coordinates $q_i$. For a quasi-one-dimensional system we may plot instead $\langle \delta x_j \rangle$ or $\langle \delta y_j \rangle$ against $\langle x_j\rangle = jL_x/N$ (see Ref. \cite{TM05}). 

For convenience, we henceforth set $L_x = 2\pi$. For the HP quasi-one-dimensional system, which is of particular interest in the present work, the vector field representations of the positive T, L and P modes respectively are found to be approximately of the form \cite{EFPZ05} \begin{align}
	\deta_T(\K) & = \sqrt{2}\begin{pmatrix} 0 \\ 1 \end{pmatrix}\cos(\K x)~,\notag\\ 	\deta_L(\K) & = \sqrt{2}\begin{pmatrix} 1 \\ 0 \end{pmatrix} \sin(\K x)~,\notag \\ 	\deta_P(\K) & = \frac{1}{\|p\|}\begin{pmatrix} p_x \\ p_y \end{pmatrix} \cos(\K x)~,\label{eqn:TLPVFR} \end{align}
for $\K= 1/2,1,3/2,\ldots$.

\subsection{Inner Product and Basis}
\label{sec:LMIPB}
The Lyapunov modes generated by the numerical scheme are orthonormal. In the vector field representation, the inner product, which defines orthonormality, is simply \begin{equation}
	\label{eqn:DIP}
	\langle \deta_i, \deta_j\rangle = \frac{1}{L_xL_y}\int_0^{L_y}\!\!\!\int_0^{L_x}\deta_i\cdot\deta_j dx dy~, \end{equation}
where $\deta_i\cdot\deta_j$ means a dot product of $\deta_i$ and $\deta_j$ in the vector field representation. Note that by conservation of momentum, under this inner product \begin{equation}
	\label{eqn:CMI}
	\int_0^{L_y}\!\!\!\int_0^{L_x}p_{x}dxdy = \int_0^{L_y}\!\!\!\int_0^{L_x}p_{y}dxdy = 0~, \end{equation}
and the T, L, P and zero modes in Eqs. (\ref{eqn:ZMVFR}) and (\ref{eqn:TLPVFR}) have unit norm under this inner product. 

>From Eq. (\ref{eqn:TLPVFR}) it follows clearly that $\langle \deta_T, \deta_L\rangle = 0$. Further, as expected $\langle \deta_a(i), \deta_a(j)\rangle = 0$, for $a = T,L$ or $P$ and half-integers or integers $i\not=j$. However, $\deta_P(\K)$ is not orthogonal to $\deta_T(\K)$ in general, since 
\begin{equation}
	\langle \deta_T(\K), \deta_P(\K)\rangle = \frac{\sqrt{2}}{4\pi\|p\|}\int_0^{2\pi}p_y\cos(2\K x) dx~. \end{equation}
Similarly the zero modes $\deta_1$ or $\deta_2$ are not strictly orthogonal to the P-modes. Of course, the T (zero) and P (P) modes must always be orthogonal, as they are Lyapunov vectors associated to different Lyapunov exponents. This difficulty may be resolved by presuming that the $p_y$ scalar field is random white noise, so that the integral is zero. 


Note also the L and P modes $\deta_L(\K)$ and $\deta_P(\K)$ are not strictly orthogonal, since \begin{equation}
	\langle \deta_L(\K), \deta_P(\K)\rangle = \frac{\sqrt{2}}{4\pi\|p\|}\int_0^{2\pi}p_x\sin(2\K x) dx~. \end{equation}
This may be easily resolved by performing an orthonormalization procedure on the L and P modes, which are associated with the same Lyapunov exponent and need not be strictly orthogonal. However, in line with our presumption for the $p_y$, we instead presume the $p_x$ are also white noise, so the integral is zero. 

Using notation in Eqs. (\ref{eqn:ZMVFR}) and (\ref{eqn:TLPVFR}), a basis of positive Lyapunov modes for the HP quasi-one-dimensional system (see Fig. \ref{fig:QDS}) is then the set \begin{equation}
	\label{eqn:LMB}
	\bigg\{\!\deta_2,\deta_3, \deta_T\Big(\frac{1}{2}\Big), \deta_L\Big(\frac{1}{2}\Big), \deta_P\Big(\frac{1}{2}\Big),\ldots, \deta_T\Big(\frac{m}{2}\Big), \deta_L\Big(\frac{m}{2}\Big), \deta_P\Big(\frac{m}{2}\Big)\!\bigg\}~, \end{equation}
where $K \equiv 3m + 2$ is the number of Lyapunov modes in the system. Note that the $\deta_1$ zero mode is absent, since translational symmetry in the $x$ direction is broken by the hard boundary conditions. 

\section{Lyapunov Mode Phenomenology}

\subsection{Numerical Dynamics}
\label{sec:LMPND}
Let $\Omega^t$ be the matrix whose columns are the $4N$ orthonormal Lyapunov vectors $\dxi^t$ generated at a particular time $t$ by the numerical scheme, with $\Omega^0 \equiv \Omega$. We call the dynamics of a Lyapunov vector or mode under the numerical scheme the numerical dynamics. This dynamics is not equivalent to the actual tangent space dynamics (\ref{eqn:TFD}), since the numerical scheme involves both evolution under the tangent flow combined with regular application of the Gram-Schmidt orthnormalization procedure. 

The numerical dynamics of the Lyapunov vectors may then be written as \begin{equation}
	\label{eqn:NDLV}
	\Omega^t = \Nd(\Omega)~,
\end{equation}
where $\Nd:T_{\xi}M \to T_{\flow{t}(\xi)}M$ is the numerical dynamics operator, which preserves orthonormality. The notation in Eq. (\ref{eqn:NDLV}) indicates operation of $\Nd$ column-wise on the matrix $\Omega$, such that if $\Omega = (\ldots|\delta\omega_j|\ldots)$, then $\Nd(\delta\omega_j) = \delta\omega_j^t$. Although $\Nd(\Omega)$ is a matrix under this notation, the operator $\Nd$ is not generally a linear map, as its form is dependent on the choice of $\Omega$ and in particular the ordering of its columns, due to the iterative nature of the Gram-Schmidt procedure. We indicate this sensitivity to the choice of Lyapunov vector basis and its ordering by the subscript $\Omega$. Note that both $T_{\xi}M$ and $T_{\flow{t}(\xi)}M \simeq \mathbb{R}^{4N}$, so that we may consider $\Nd:\mathbb{R}^{4N} \to \mathbb{R}^{4N}$, or $\Nd: \mathbb{R}^2 \to \mathbb{R}^2$ in the vector field representation. 

The Lyapunov modes generated by the numerical scheme may exhibit time-periodic behavior. That is, for a Lyapunov mode $\deta^t$ at some time $t$, there exists some period $\tau$ such that $\deta^t = \deta^{\tau +t}$. The longitudinal modes generally exhibit the most obvious time-periodicity, as can be seen for the HP quasi-one-dimensional system in Figs. 4 and 5 of \cite{TM05}. 

Many previous works (e.g. \cite{TM05,EFPZ05}) have noted that the numerical dynamics of an LP mode pair are well-approximated by a time-modulated linear combinations of their corresponding L and P modes at some fixed time, say $t_0=0$. For example, putting $\tau = 2\pi$, we might have \begin{equation}
	\label{eqn:LCLPM}
	\deta^t \equiv \begin{pmatrix} \deta^t_x \\ \deta^t_y \end{pmatrix} \approx \sqrt{2}\cos(t)\begin{pmatrix}1 \\ 0\end{pmatrix}\sin(x/2) + 
\sin(t) \frac{1}{\|p\|}\begin{pmatrix} p_x \\ p_y \end{pmatrix}\cos(x/2)~, \end{equation}
as an approximation to longitudinal mode 197 shown in Figs. 4 and 5 of Ref. \cite{TM05}. 

Figure \ref{fig:LMFO} shows the contour plots for the predicted LP mode numerical dynamics according to Eq. (\ref{eqn:LCLPM}). Note that we neglect terms involving $p_x$ in the $x$-component of the mode, since we assumed this momentum component was random white noise. Similarly, the $y$-component of an LP mode includes only terms with $p_y$ coefficients, so we plot instead $\deta^t_y/p_y$. We will plot all other modes in this way. 

\begin{figure}[ht]
	\begin{center}
	\includegraphics{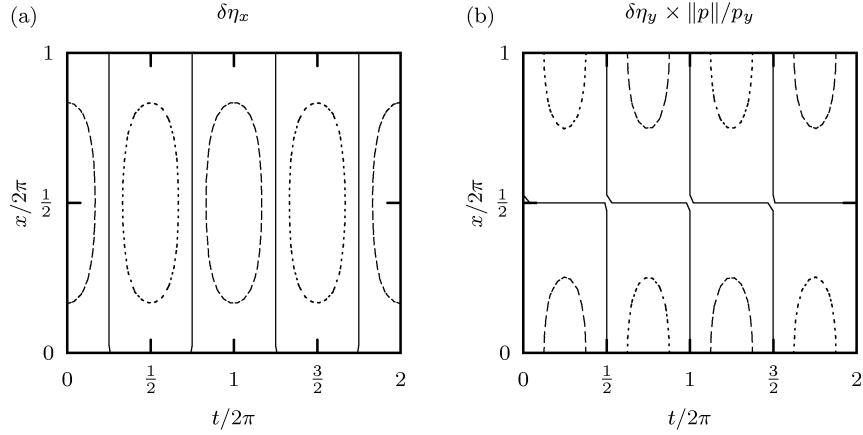}
	\end{center}
	\caption{Contour plots of the predicted LP mode numerical dynamics in the HP quasi-one-dimensional system according to Eq. (\ref{eqn:LCLPM}) for (a) $\deta_x^t$ and (b) $\deta_y^t/p_y$. Three contours are shown in each plot: a contour at the half maximum height level, $1/\sqrt{2}$ (dashed line); a contour at the zero level (solid line); and a contour at the half minimum height level, $-1/\sqrt{2}$ (dotted line).} 	\label{fig:LMFO}
\end{figure}

The empirical contour plots in Fig. 5 of Ref. \cite{TM05} reveal notable features of the LP mode dynamics such as triangular-like half-height contours that have a dimple on their short side. These are not reproduced in Fig. \ref{fig:LMFO} here. Hence, at least in the quasi-one-dimensional case, it is clear that the time-periodic behavior of the LP mode is not fully described by Eq. (\ref{eqn:LCLPM}). There are therefore two questions which arise concerning the numerical dynamics: can a better description of the numerical dynamics be found; and moreover is the time periodicity merely an artifact of the numerical scheme or a more fundamental property of the tangent dynamics. 

\subsection{Separation of the Numerical Dynamics} 

The numerical dynamics operator $\Nd$ is bijective, so that at some time $t$ a Lyapunov vector may be described as some linear combination of all the other Lyapunov vectors at some initial time $t = 0$. However, the partially successful description (\ref{eqn:LCLPM}) of an LP mode at any time $t$ involves a linear combination of only its corresponding $L$ and $P$ modes at $t=0$. Validity of Eq. (\ref{eqn:LCLPM}) implies then that the numerical dynamics of this LP mode pair separates from the dynamics of other Lyapunov modes or vectors, and that the numerical dynamics is a linear map. We now use this general principle to obtain a better description of the numerical dynamics with the following claim. 

Let $\{\deta_k\}$, $k= 1,\ldots,K$ be a basis of positive Lyapunov modes ordered as in Eq. (\ref{eqn:LMB}). Let $\Xi^t$ be the $2\times K$ matrix with orthonormal columns whose $k$th column is the positive mode $\deta^t_k$ in the vector field representation. For the HP quasi-one-dimensional system, by Eq. (\ref{eqn:LMB}) we set the vector field representation form for $\Xi^0 \equiv \Xi$ to be the $2\times11$ matrix \begin{equation}
	\label{eqn:XM}
	\Xi = \bigg(\deta_2\Big|\deta_3\Big| \deta_T\Big(\frac{1}{2}\Big)\Big| \deta_L\Big(\frac{1}{2}\Big)\Big| \deta_P\Big(\frac{1}{2}\Big)\Big| \ldots\Big| \deta_T\Big(\frac{3}{2}\Big)\Big| \deta_L\Big(\frac{3}{2}\Big)\Big| \deta_P\Big(\frac{3}{2}\Big)\bigg)~. \end{equation}

\begin{claim}
After some time interval $t$, the positive Lyapunov modes $\deta^t_k$ are time-modulated linear combinations of only the positive initial modes $\deta_k$ such that \begin{equation}
	\label{eqn:CND}
	\Xi^t =\Xi \A(t)~,
\end{equation}
where $\A(t)$ is a modal, orthogonal $K\times K$ matrix. We call $\A(t)$ the admixture matrix since it represents linear combinations of Lyapunov modes. We further claim the longitudinal (transverse) modes are linear combinations only of the initial longitudinal (transverse) modes. \end{claim}

According to Eq. \ref{eqn:CND}, the $j$th column of $\A(t)$, $a_j(t)$, is matrix-multiplied with the matrix $\Xi$ to produce the $j$th mode at any time $t$. We say that $a_j(t)$ generates the $j$th mode, and note that $\A(t)$ must be orthogonal since the numerical scheme produces orthonormal sets of Lyapunov modes: both $\Xi^t$ and $\Xi$ have orthonormal columns, and hence $\A(t)$ must be orthogonal. The claim (\ref{eqn:CND}) implies that the numerical dynamics of either the longitudinal or transverse modes separates from that of the rest of the tangent space. 

\subsection{Admixture Matrix}
In order to test the claim (\ref{eqn:CND}), we now seek to construct the matrix $\A(t)$ for the HP quasi-one-dimensional system (see Fig. \ref{fig:QDS}). For simplicity, we fix the period $\tau = 2\pi$, and we only construct explicitly the columns of $\A(t)$, which generate the zero modes and those modes corresponding to smallest wavenumber $k_x = 1/2$. That is we consider only the first five columns of $\A(t)$. 

The positive zero modes are static under the numerical dynamics (see App. \ref{sec:TFSZM}), and assuming the transverse modes are also static\footnote{Fourier analysis of the empirical transverse mode numerical dynamics is required to confirm this. We do not address this issue in the present work.}, from Eq.~(\ref{eqn:LCLPM}) an initial guess at the matrix $\A(t)$ is the block diagonal matrix \begin{equation}
\label{eqn:IAAM}
\A(t)\equiv \Big(a_1(t)\Big|\ldots\Big|a_{11}(t)\Big) = \begin{pmatrix} 1 	& 0 & & 0	& & \ldots\\
0 	& 1 \\
	& & 1 & 0	& 0\\
0 	& & 0 & \cos(t) & \sin(t) \\
	& & 0 & -\sin(t) & \cos(t)\\
\vdots&	& &	&	&\ddots
\end{pmatrix}~,
\end{equation}
where the submatrices on the diagonal are rotation matrices with period $2\pi$. 

Now, the desired finer details of the LP modes, such as dimples in the half-height contours, appear to occur at double the frequency of the actual mode, so we presume then that there are terms such as $\cos(t)\cos(2t)$ in the admixture matrix: we expect a modulation on top of the modulation described in Eq. (\ref{eqn:LCLPM}), so that the former has double the latter's frequency. Since $\A(t)$ is orthogonal, its columns must have unit norm, so if $\cos(t)\cos(2t)$ occurs in a column then so must $\cos(t)\sin(2t)$, $\sin(t)\cos(2t)$ and $\sin(t)\sin(2t)$. Treating Eq. (\ref{eqn:IAAM}) as a starting approximation, and enforcing orthonormality, a possible choice of the fourth and fifth columns of $\A(t)$ is \begin{align}
	a_4(t) & = \frac{1}{\sqrt{1+\alpha^2}}\Big(0,0,0,\notag\\ 	& \quad\quad\quad\C(t),-\s(t),0,\alpha\C(t)\C(2t),-\alpha\s(t)\s(2t),0,\alpha\s(t)\C(2t),-\alpha\C(t)\s(2t)\Big)\label{eqn:A4}\\ 	a_5(t) & = \frac{1}{\sqrt{1+\alpha^2}}\Big(0,0,0,\notag\\ 	& \quad\quad\quad\s(t),\C(t),0,\alpha\s(t)\C(2t),-\alpha\C(t)\s(2t),0,-\alpha\C(t)\C(2t),\alpha\s(t)\s(2t)\Big)\label{eqn:A5} \end{align}
where $\C(t) \equiv \cos(t)$, $\s(t) \equiv \sin(t)$, and $\alpha > 0$ is some small real number. 

Plots of the predicted modes generated by these columns are shown in Fig. \ref{fig:LM}. As can be seen by comparison with Fig. 5 in Ref. \cite{TM05}, the finer structure of the mode dynamics, such as the triangular positive and negative half-height contours with dimples, appear to be properly reproduced and oriented by Eqs. (\ref{eqn:A4}) and (\ref{eqn:A5}). Further, the zero level contour of the $\deta_y/p_y$ components appears to match closely the numerical data, although this is more difficult to determine due to numerical noise. Equations (\ref{eqn:A4}) and (\ref{eqn:A5}) therefore appear to better describe the numerical dynamics than Eq. (\ref{eqn:LCLPM}). 

\begin{figure}[ht]
	\begin{center}	
	\includegraphics[scale=0.99]{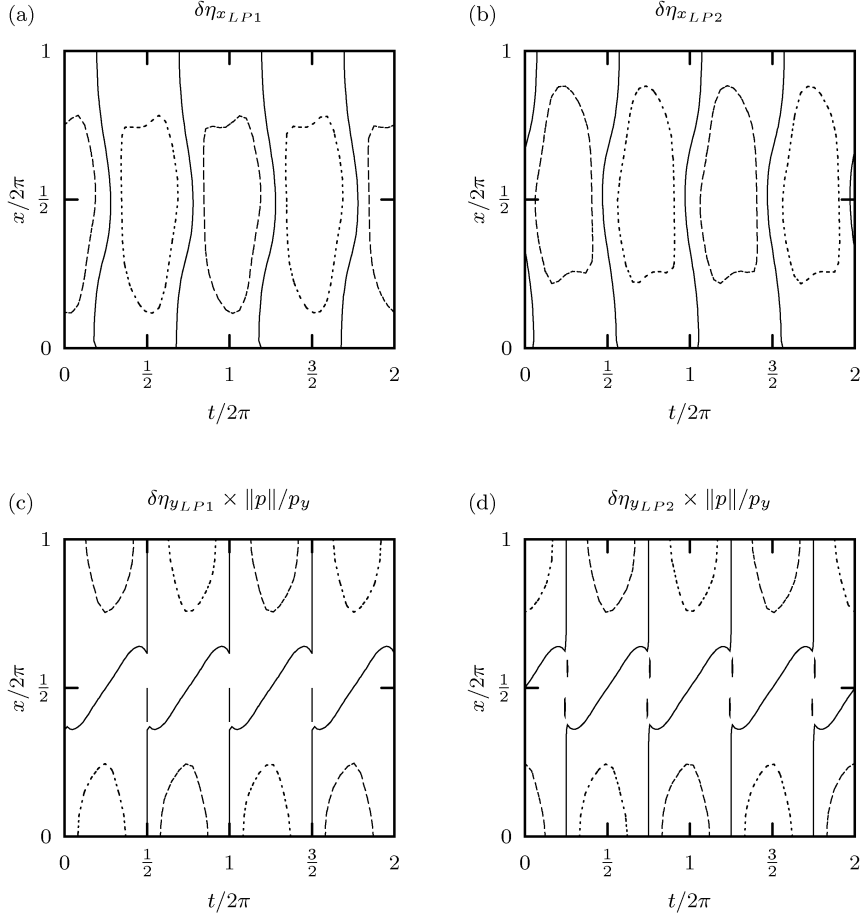} 	\end{center}
	\caption{Contour plots of the predicted first and second LP mode numerical dynamics in the HP quasi-one-dimensional system according to Eqs. (\ref{eqn:A4}) and (\ref{eqn:A5}) for (a) $\deta_x$ of the first LP mode, (b) $\deta_x$ of the second LP mode, (c) $\deta_y/p_y$ of the first LP mode, and (d) $\deta_y/p_y$ of the second LP mode. The parameter $\alpha = 0.2$. Three contours are shown in each plot: a contour at the half maximum height level, $1/\sqrt{2}$ (dashed line); a contour at the zero level (solid line); and a contour at the half minimum height level, $-1/\sqrt{2}$ (dotted line). Compare this figure with Fig. 5 of Ref. \cite{TM05}.} 	\label{fig:LM}
\end{figure}

\section{Lyapunov Mode Dynamics}
\label{sec:LMD}

\subsection{Full Numerical Dynamics of the Lyapunov Modes} \label{sec:LMDFNDLM}
For a basis of positive Lyapunov modes $\{\deta_k\}$, $k=1,\ldots,K$ as in Eq. (\ref{eqn:LMB}), we may define a corresponding basis of negative modes (see App. \ref{sec:TFFHP}) by putting \begin{equation}
	\label{eqn:NLMB}
	\deta_{k+K} = J_{4N}\deta_k~.
\end{equation}
Here for some integer $L$, $J_{2L}$ is the $2L\times 2L$ matrix \begin{equation}
	\label{eqn:DSMJ}
	J_{2L} = \begin{pmatrix} 0 & -I \\ I & 0 \end{pmatrix}~, \end{equation}
in which $0$ ($I$) is the $L\times L$ null (identity) matrix. 

We define the matrix $\Xi = (\ldots| \deta_k |\ldots)$ similarly to Eq. (\ref{eqn:XM}), and then by Eq. (\ref{eqn:NLMB}), $\XiJ = (\deta_k)_{k=1}^{2K}$ is a matrix whose columns form an orthonormal basis of all the Lyapunov modes. Note that we consider here $\deta_k \in \mathbb{R}^{4N}$, rather than the vector field representation, and the inner product of modes is then simply the dot product $\deta_i\cdot\deta_j \equiv \deta_i^T\deta_j$, rather than an integral (\ref{eqn:DIP}). The following arguments will, of course, hold for the vector field representation and its inner product. Note also that the orthonormality of the columns of $\XiJ$ is due to the property \begin{equation}
\Xi^TJ_{4N}\Xi = 0.
\end{equation}
This depends on the fact that positive mode $\deta_k \in \Ur{j}{\xi}$, for some $j \le (l+1)/2$. However, $J\deta_k \in J\Ur{j}{\xi} = \Ur{l-j+1}{\xi}$ (see App. \ref{sec:TFFHP}) is a negative mode and the spaces $\Ur{j}{\xi}$ are all orthogonal, so $\deta_i \cdot J\deta_k = 0$ for any $i$ and $k$. 

The claim (\ref{eqn:CND}) describes the numerical dynamics of only the positive modes. However, the numerical dynamics $\Nd$ (\ref{eqn:NDLV}) is a symplectic operator (see Appendix \ref{sec:ASND}) such that $(J_{4N}\Xi)^t = J_{4N}(\Xi)^t$. Hence by the claim (\ref{eqn:CND}), the negative modes must have dynamics $(J_{4N}\Xi)^t = J_{4N}(\Xi)^t = J_{4N}\Xi\A(t)$. The claim (\ref{eqn:CND}) therefore implies that the full numerical dynamics of all the modes is simply \begin{equation}
	\label{eqn:NDB}
	\BXiJ^t = \BXiJ\N(t),\ \mbox{where}\ \N(t) \equiv \begin{pmatrix}\A(t) & 0\\ 0 & \A(t) \end{pmatrix}~. \end{equation}

Since, according to the claim (\ref{eqn:CND}), the matrix $\A(t)$ is orthogonal, then so is the $2K\times 2K$ matrix $\N(t)$. The orthogonality of $\A(t)$ also means that $\N(t)$ is a symplectic matrix. That is, \begin{equation}
	\label{eqn:SMN}
	\N(t)^TJ_{2K}\N(t) = J_{2K}~,
\end{equation}
as expected (see App. \ref{sec:ASND}).

Further, by Eq. (\ref{eqn:NDLV}), the numerical dynamics of the modes may be generally written as \begin{equation}
	\label{eqn:GND}
	\BXiJ^t = \Nd\bigg[\BXiJ\bigg]~,
\end{equation}
where here $\Nd:\mathbb{R}^{4N} \to \mathbb{R}^{4N}$. Note that the ordering of $\XiJ$ does not matter as $\Nd$ is a column-wise operator (see Eq. \ref{eqn:NDLV}). Let $T_{\xi}X \equiv \mbox{span}\{\deta_k\}_{k=1}^{2K}$. Then the numerical dynamics (\ref{eqn:NDB}) implies that $T_{\xi}X$ is an invariant subspace of $\Nd$, that is $\Nd(T_{\xi}X) = T_{\xi}X$, or in other words, the numerical dynamics on $T_{\xi}X$ separates from the dynamics on the rest of the tangent space. Furthermore, comparing Eq. (\ref{eqn:NDB}) with Eq. (\ref{eqn:GND}) it follows that $\Nd[\XiJ] = \XiJ\N(t)$, and hence \begin{equation}
	\label{eqn:SNDTX}
	\Nd(\deta_k) = \BXiJ\N(t)\BXiJ^T\deta_k~. \end{equation}
Since $\deta_k$ is a basis of $T_{\xi}X$, then it must be that the numerical dynamics $\Nd$ is a linear operator on $T_{\xi}X$, provided the claim (\ref{eqn:CND}) is true. 

\subsection{Connection to the Tangent Dynamics} Consider the usual tangent dynamics (\ref{eqn:TFD}) after normalization, which we henceforth call the normalized tangent dynamics. In general, the normalized tangent dynamics of the modes is a linear map and may be written as \begin{equation}
	\label{eqn:DNTD}
	D(t)\BXiJ = \tanflow{\xi}{t}\BXiJ\mbox{diag}\bigg\{\frac{1}{\|\tanflow{\xi}{t}\deta_k\|}\bigg\}_{k=1}^{2K}~. \end{equation}
We define $\deta_k(t) = D(t)\deta_k$.

Suppose the normalized tangent dynamics on $T_{\xi}X$ separates from the other dynamics, so that $D(t)\deta_k = \XiJ\D(t)\XiJ^T\deta_k$ for some matrix $\D(t)$ similarly to Eq. (\ref{eqn:SNDTX}). Then, since the $\deta_k$ form a basis of $T_{\xi}X$, it follows that we have \begin{equation}
	\label{eqn:SNTD}
	D(t)\BXiJ = \BXiJ\D(t)
\end{equation}
noting orthonormality of the columns of $\XiJ$ means 
\begin{equation}
	\label{eqn:MMO}
	\BXiJ^T\BXiJ = I_{2K}~,
\end{equation}
where $I_{2K}$ is the $2K\times2K$ identity matrix 

The normalized tangent dynamics of the zero modes clearly separates from that of the rest of the tangent space (see App. \ref{sec:TFSZM}). From Eqs. (\ref{eqn:DSZM}), (\ref{eqn:DPZM}) and (\ref{eqn:DEZM}) we have that \begin{equation}
	D(t)\Big(\deta_1\Big|\ldots\Big|\deta_{2s+2}\Big) = \Big(\deta_1\Big|\ldots\Big|\deta_{2s+2}\Big)\D_0(t) \end{equation}
where $\deta_1,\ldots,\deta_{2s+2}$ are the zero modes for a system with $s$ spatial symmetries and \begin{equation}
	\D_0(t) = \begin{pmatrix}
			I_{s+1} & \beta tI_{s+1}[(\beta t)^2 +1]^{-1/2}\\ 			0 & I_{s+1}[(\beta t)^2 +1]^{-1/2}
			\end{pmatrix}~.
\end{equation}
We may trivially decompose $\D_0(t)$ as follows: \begin{equation}
	\label{eqn:QRZM}
	\D_0(t) = I_{2s+2}
			\begin{pmatrix}
			I_{s+1} & \beta tI_{s+1}[(\beta t)^2 +1]^{-1/2}\\ 			0 & I_{s+1}[(\beta t)^2 +1]^{-1/2}
			\end{pmatrix}~,
\end{equation}	
noting that the identity $I_{2s+2}$ is simply the numerical dynamics of the zero modes $\N_0(t)$. Since the Lyapunov modes appear to be modulations of the zero modes \cite{EFPZ05}, we then extend this dynamical form and presume that the normalized tangent dynamics of the modes is \begin{equation}
\label{eqn:QRM}
	\D(t) = \begin{pmatrix} \A(t) & 0 \\ 0 & \A(t) \end{pmatrix} 		\begin{pmatrix}
		I_K & \beta tI_K[(\beta t)^2 +1]^{-1/2}\\ 		0 & I_K[(\beta t)^2 +1]^{-1/2}
		\end{pmatrix}~,
\end{equation}
where we have replaced the numerical dynamics of the zero modes $\N_0(t)$ by our claimed numerical dynamics $\N(t)$ (\ref{eqn:NDB}) for the modes, and adjusted the dimensionality of the matrix on the far right accordingly. That is we claim \begin{equation}
\label{eqn:PFNTD}
	\D(t) = \begin{pmatrix}
			\A(t) & \beta t\A(t)[(\beta t)^2 +1]^{-1/2}\\ 			0 & \A(t)[(\beta t)^2 +1]^{-1/2}
			\end{pmatrix}~,
\end{equation}
is a general form for the normalized tangent dynamics, where $\A(t)$ is some periodic, orthogonal matrix. Note that Eqs. (\ref{eqn:QRZM}) and (\ref{eqn:QRM}) are the QR factorizations of $\D_0(t)$ and $\D(t)$ respectively. This is expected since the numerical scheme differs from the normalized tangent dynamics only by the application of the Gram-Schmidt orthonormalization procedure. 

\subsection{Symplecticity of the Normalized Tangent Dynamics} 

We may further re-write Eq. (\ref{eqn:QRM}) as $\D(t) = \N(t)\mathcal{R}(t)$, defining $\mathcal{R}(t)$ in the obvious way. One finds that the matrix $\mathcal{R}(t)$ is `almost symplectic', in the sense that $\mathcal{R}(t)^TJ_{2K}\mathcal{R}(t) = J_{2K}[(\beta t)^2 +1]^{-1/2}$. By Eq. (\ref{eqn:SMN}) it follows that $\D(t)$ is similarly almost symplectic, i.e. \begin{equation}
	\label{eqn:ASNTD}
	\D(t)^TJ_{2K}\D(t) = J_{2K}[(\beta t)^2 +1]^{-1/2}~. \end{equation}

Now, the matrix $\XiJ$ has the property that \begin{equation}
	\label{eqn:MMS}
	\BXiJ^TJ_{4N}\BXiJ = J_{2K},
\end{equation}
and from Eqs. (\ref{eqn:DNTD}), (\ref{eqn:SNTD}) and (\ref{eqn:MMO}) we have explictly that \begin{equation}
	\label{eqn:EFNTD}
	\D(t) = \BXiJ^T\tanflow{\xi}{t}\BXiJ\mbox{diag}\bigg\{\frac{1}{\|\tanflow{\xi}{t}\deta_k\|}\bigg\}_{k=1}^{2K}~. \end{equation}
Separability of the normalized tangent dynamics (\ref{eqn:SNTD}) implies that its projection onto $T_{\xi}X$, which is $\XiJ\XiJ^T\tanflow{\xi}{t}\XiJ$, is simply equal to $\tanflow{\xi}{t}\XiJ$. Hence from this together with Eqs. (\ref{eqn:EFNTD}) and (\ref{eqn:MMS}) we have that Eq. (\ref{eqn:ASNTD}) is equivalent to \begin{equation}		\mbox{diag}\bigg\{\frac{1}{\|\tanflow{\xi}{t}\deta_k\|}\bigg\}_{k=1}^{2K}J_{2K}\mbox{diag}\bigg\{\frac{1}{\|\tanflow{\xi}{t}\deta_k\|}\bigg\}_{k=1}^{2K} = J_{2K}[(\beta t)^2 +1]^{-1/2}~, \end{equation}
which in turn holds if and only if
\begin{equation}
	\label{eqn:MGLM}
	\log\|\tanflow{\xi}{t}\deta_k\| + \log\|\tanflow{\xi}{t}J_{4N}\deta_k\| = \log\big\{[(\beta t)^2 +1]^{1/2}\big\} \end{equation}
for $k = 1,\ldots,K$.

\subsection{Lyapunov Mode Stability}
\label{sec:LMDLMS}
The empirical multiplicity of the zero modes implies (see App. \ref{sec:TFEZS}) that the spaces of Lyapunov zero modes satisfy \begin{equation}
\Ur{m}{\xi} = \W{m}{\xi} = \Uf{m}{\xi}
\end{equation}
where $m = (l+1)/2$ and $\W{m}{\xi}$ is the zero Oseledec space. 

Suppose that the symplecticity property (\ref{eqn:MGLM}) of the Lyapunov modes holds. We call the spaces $\Ur{j}{\xi}$, whose elements are Lyapunov modes, mode spaces, and the symplecticity property then implies that for all $\dxi \in \Ur{j}{\xi}$, a mode space, \begin{equation}
	\label{eqn:SRLM}
	\lim_{t \to \pm\infty}\frac{1}{|t|}\|\tanflow{\xi}{t}\dxi\| + \lim_{t \to \pm\infty}\frac{1}{|t|}\|\tanflow{\xi}{t}J\dxi\| = 0 \end{equation}
since $\log[(\beta t)^2 +1]/|t| \to 0 $ as $t \to \pm\infty$. Certainly Eq. (\ref{eqn:SRLM}) is true for the limit $t \to -\infty$ by the general theory of tangent vectors, as $\deta_k$ is the Lyapunov vector associated to Lyapunov exponent $\lambda_k$ and $\deta_{k+K}$ to $-\lambda_k$ in the negative time limit (see Apps. \ref{sec:TFTSD}, \ref{sec:TFHP}). However, this result is not necessarily true for $t \to +\infty$, and thus is a particular property of the form of our claimed normalized tangent dynamics (\ref{eqn:PFNTD}). We now show that Eq. (\ref{eqn:SRLM}) means that \begin{equation}
	\label{eqn:ERFEO}
	\Ur{j}{\xi} = \W{j}{\xi},
\end{equation}
where $j$ is such that $\Ur{j}{\xi}$ is a mode space. 

In order to show $\Ur{j}{\xi} = \W{j}{\xi}$, it is sufficient to show $\Ur{j}{\xi} \subseteq \W{j}{\xi}$ because $\dim \W{m}{\xi} = \dim\Ur{m}{\xi}$ (see App. \ref{sec:TFCS}). First of all, it follows from Eq. (\ref{eqn:RVP2}) that \begin{equation}
	\label{eqn:LMI1}
	\Ur{j}{\xi} \subseteq \bigg\{\dxi \in T_{\xi}M : \lim_{t \to +\infty}\frac{1}{|t|}\log\| \tanflow{\xi}{t}\dxi \| \ge \lam{j}\bigg\}\cup\{\bm{0}\}~. \end{equation}
We then have that, by Eqs. (\ref{eqn:HRLE}) and (\ref{eqn:HRES}), \begin{align}
	\Ur{j}{\xi}
	& = J\Ur{l-j+1}{\xi} \notag\\
	& \subseteq \bigg\{\dxi \in T_{\xi}M : \lim_{t \to +\infty}\frac{1}{|t|}\log\| \tanflow{\xi}{t}J\dxi \| \ge -\lam{j}\bigg\}\cup\{\bm{0}\}\notag\\ 	& = \bigg\{\dxi \in T_{\xi}M : \lim_{t \to +\infty}\frac{1}{|t|}\log\| \tanflow{\xi}{t}\dxi \| \le \lam{j}\bigg\}\cup\{\bm{0}\}~, \label{eqn:LMI2} \end{align}
the last step of which follows from the symplecticity relation (\ref{eqn:SRLM}). Comparing Eqs. (\ref{eqn:LMI1}) and (\ref{eqn:LMI2}) produces \begin{equation}
	\Ur{j}{\xi} \subseteq \bigg\{\dxi \in T_{\xi}M : \lim_{t \to +\infty}\frac{1}{|t|}\log\| \tanflow{\xi}{t}\dxi \| = \lam{j}\bigg\}\cup\{\bm{0}\}~, \end{equation}
for a mode space $\Ur{j}{\xi}$.
But by Eqs. (\ref{eqn:RPTS}) and (\ref{eqn:RELVS}), for any eigenspace it is the case that \begin{equation}
	\Ur{j}{\xi} \subseteq \bigg\{\dxi \in T_{\xi}M : \lim_{t \to -\infty}\frac{1}{|t|}\log\| \tanflow{\xi}{t}\dxi \| = -\lam{j}\bigg\}\cup\{\bm{0}\}~, \end{equation}
and thus
\begin{equation}
	\Ur{j}{\xi} \subseteq \bigg\{\dxi \in T_{\xi}M : \lim_{t \to \pm\infty}\frac{1}{|t|}\log\| \tanflow{\xi}{t}\dxi \| = \pm\lam{j}\bigg\}\cup\{\bm{0}\} \equiv \W{j}{\xi}~. \end{equation}

Note that the result (\ref{eqn:ERFEO}) is a partial extension of the corresponding result for the zero eigenspaces (\ref{eqn:EZS}). Moreover, Eq. (\ref{eqn:ERFEO}) implies that the mode spaces are covariant (see App. {\ref{sec:TFCS}) and the Lyapunov modes have well-defined Lyapunov exponents in the positive time limit. Hence our claimed tangent dynamics, via its symplecticity property (\ref{eqn:MGLM}), implies that the modes are phase space perturbations which also have their (in)stability in the positive time limit characterized by the same Lyapunov exponent as in the negative limit. They therefore attain additional physical significance. 


\section{Numerical Results}

\subsection{Modified Numerical Scheme}
We now check the validity of the claimed tangent dynamics (\ref{eqn:PFNTD}) via the following modification of the numerical scheme. Firstly, we choose $2N$ linearly independent tangent vectors, and allow them to converge to an orthonormal set of Lyapunov vectors under the usual numerical scheme. These approximate Lyapunov vectors, which we denote as $\dxi_j$, are a set of positive Lyapunov vectors at some initial time $t_0$, and we set the negative Lyapunov vectors to be \begin{equation}
\dxi_{4N-j+1} = J_{4N}\dxi_j~.
\end{equation}
We call $\dxi_j$ and $\dxi_{4N-j+1}$ for some $j$ a symplectic pair of Lyapunov vectors. The ordering of the Lyapunov vectors here corresponds to the order in which the numerical scheme (see Sec. \ref{sec:LVMNS}) produces the Lyapunov vectors: most unstable Lyapunov vector to most stable Lyapunov vector in the positive time limit. This ordering is a convenient choice for the purpose of presenting numerical data, but differs from the ordering of the positive Lyapunov vectors $\dxi_j$, $j=1,\ldots,2N$, implied in Eq. (\ref{eqn:LMB}). However, the ordering of the negative Lyapunov modes here is the same as that adopted in Eq. (\ref{eqn:NLMB}). 

Let $\Omega = (\ldots|\dxi_j|\ldots)$ for $j = 1,\ldots, 4N$, and put the initial time $t_0 = 0$. The middle $2K$ columns of $\Omega$ are approximately the initial Lyapunov modes. We denote these modes as $\deta_j$, with $j = 2N-K+1,\ldots,2N+K$ to match the indexing of the $\dxi_j$. 

We wish to consider the evolution of the $\deta_j$ under the normalized tangent dynamics only, so that the Gram-Schmidt orthonormalization, which occurs at regular intervals, is replaced by just normalization. Unfortunately, the chaotic nature of the system means that the numerics is inherently unstable without the Gram-Schmidt procedure, because numerical noise introduces components of the strongly dynamically localized Lyapunov vectors, which exponentially dominate the modes under further evolution. To reduce this numerical instability, we therefore continue to Gram-Schmidt orthonormalize the modes with respect to the first $2N-K$ Lyapunov vectors. This modified numerical scheme for the modes may be written explicitly as \begin{equation}
	\label{eqn:MNTD}
	\deta_j(t) = C_k\Big(I_{4N} - \sum_{i=1}^{2N-K}\!\!\dxi_i^t(\dxi_i^t)^T\Big)\tanflow{\xi}{t}\deta_j \end{equation}
for $j=2N-K+1,\ldots,2N+K$, where $C_k$ is the appropriate normalization constant. 

We compute the Lyapunov modes $\deta_j(t)$ for a $N=50$ particle quasi-one-dimensional system with periodic boundary conditions and density $0.80$. In all the following computations we use an initial set of positive Lyapunov vectors $\dxi_j$, $j=1,\ldots,100$, generated after $1200$ collisions per particle, with the usual Gram-Schmidt procedure applied after each collision. 

In Fig. \ref{fig:LS} we show the Lyapunov spectrum for this system. The zero exponents and first three positive steps are readily apparent: the two-point steps consist of a two-fold degenerate Lyapunov exponent associated with transverse modes, whilst the single four-point step consists of a four-fold degenerate Lyapunov exponent associated with longitudinal modes. We label these steps 0, T1, LP1, and T2 according to increasing Lyapunov exponent, as shown in Fig. \ref{fig:LS}. Using the Lyapunov exponent value of these steps to calculate the constants in the dispersion relation (\ref{eqn:LEDR}), one identifies a further two-point step, T3, so we therefore set the number of positive modes to be $K=13$. 

\begin{figure}[ht]
\begin{center}
	\includegraphics{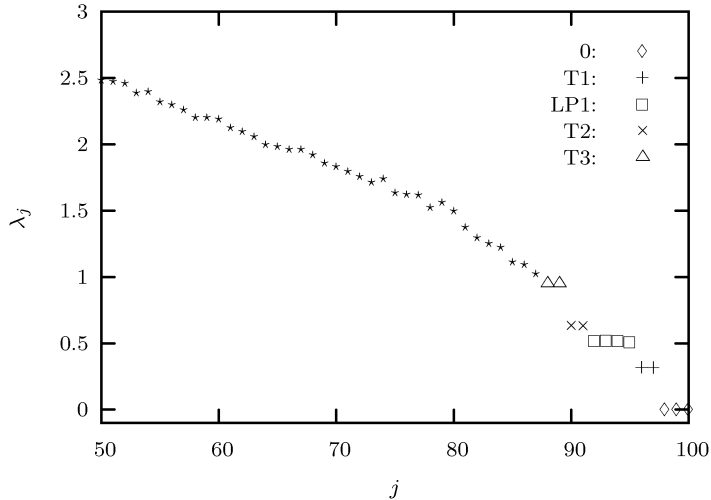}
\end{center}
	\caption{Partial Lyapunov spectrum for a $50$ particle quasi-one-dimensional system with periodic boundary conditions. Lyapunov exponents $\lambda_j$ are indexed by $j$ according to their order of calculation by the numerical scheme. Only the positive exponents are shown, and for the sake of the clarity of the Lyapunov steps, only the second fifty Lyapunov exponents are plotted. Lyapunov exponents lying on identified Lyapunov steps are indicated by different symbols with the corresponding step label shown in the legend.} 	\label{fig:LS}
\end{figure}

\subsection{Normed Differences}
\label{sec:NRND}
>From Eq. (\ref{eqn:PFNTD}), for index $j = 88,\ldots,100$ one finds that the positive Lyapunov modes have normalized tangent dynamics $\deta_{j}(t) = \Xi a_{100-j+1}(t)$, while the negative modes have 
\begin{equation}
	\label{eqn:NTDNM}
	\deta_{200-j+1}(t) = \Xi(It +J_{200})a_{100-j+1}(t)/[(\beta t)^2 +1]^{-1/2}~, \end{equation}
where $a_j(t)$ is the $j$th column of $\A(t)$. Comparison of this normalized tangent dynamics with the numerical dynamics (\ref{eqn:NDB}) reveals that the positive modes, $\deta_j(t)$, should coincide with $\deta_j^t$, where as usual $\deta_j^t$ denotes numerical dynamics. That is, for $j =88,\ldots,100$, we should have that the normed difference \begin{equation}
	\label{eqn:NDPM}
	\|\deta_j(t) - \deta_j^t\|^2 = 0.
\end{equation}
Note that if this condition is satisfied, then the normalized tangent dynamics must preserve orthonormality on the positive modes \footnote{The converse of this does not necessarily hold.}, as predicted in Eq. (\ref{eqn:PFNTD}), so we obtain a measure of the orthogonality of $\A(t)$. According to Eq. (\ref{eqn:NDB}), $(J_{200}\deta_j)^t = J_{200}\Xi a_{100-j+1}(t)$ and hence from Eq. (\ref{eqn:NTDNM}) it follows that we additionally expect \begin{equation}
	\label{eqn:NDNM}
	\|\deta_j(t) - \deta_j^t\|^2 = 2\bigg(1 -\frac{1}{[(\beta t)^2 +1]^{1/2}}\bigg) \end{equation}
for the negative modes.

Figure \ref{fig:ND}a shows the norm difference $\|\deta_j(t) - \deta_j^t\|^2$ for the positive modes. The zero modes' normed differences remain precisely zero, and Eq. (\ref{eqn:NDPM}) seems to be well-satisfied for the T1 positive modes. However the LP1 positive modes exhibit growth in their respective normed differences. 

\begin{figure}[tp]
	\begin{center}
	\includegraphics{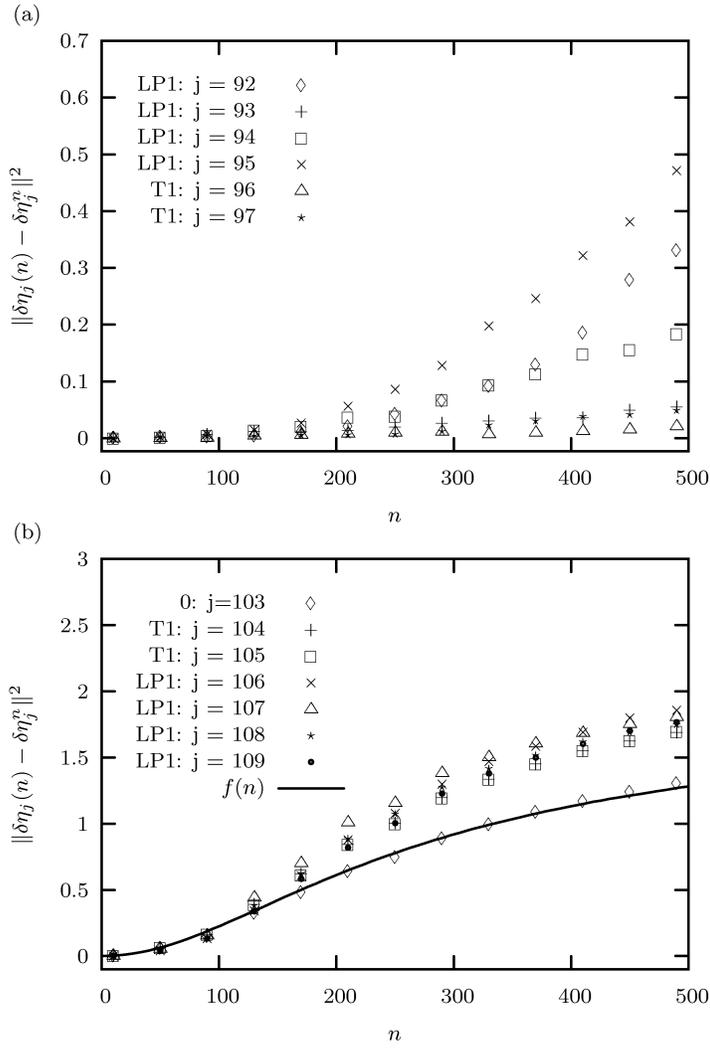}
	\end{center}
	\caption{Normed differences $\|\deta_j(n) - \deta_j^n\|^2$ for positive (a) and negative (b) modes over $500$ collisions, with $j = 92,\ldots,97$ and $j=104,\ldots,109$ respectively. $n$ is the collision number. The corresponding Lyapunov step label for each mode is shown in the legend. (a): The positive zero modes are not shown, but all satisfy $\|\deta_j(n) - \deta^n_j\| = 0$ within machine precision. (b): The function $f(n) = 2[1-1/\sqrt{(\beta n)^2 +1}]$ is best fitted to the negative zero modes at $\beta = 1/192.7$ with reduced $\chi^2 = 3.7\times10^{-4}$.} 	\label{fig:ND}
\end{figure}

It is instructive to compare this growth rate with that of the non-modes. We therefore compute $\|\dxi_j(t) - \dxi_j^t\|^2$ for $j = 62,\ldots,67$ in Fig. \ref{fig:NDNM}, where $\dxi_j(t)$ denotes evolution of the non-mode $\dxi_j$ under the modified dynamics (\ref{eqn:MNTD}) with $K = 40$. The non-modes exhibit much faster, and more erratic growth in their normed differences than the Lyapunov modes. This suggests that we may consider the growth of the modes' normed differences to be small. Further, rotation toward the more unstable modes, due to numerical instability, may be responsible for the small growth in the LP1 modes' normed differences (see Sec. \ref{sec:NRIP}), so a reasonable conclusion seems to be that Eq. (\ref{eqn:NDPM}) may hold up to numerical instability. 

\begin{figure}[htp]
	\begin{center}
	\includegraphics{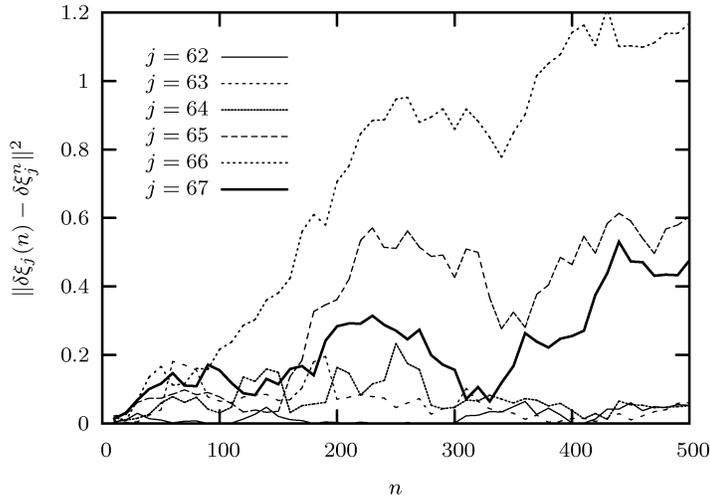}
	\end{center}
	\caption{Normed differences $\|\dxi_j(n) - \dxi_j^n\|^2$ for non-modes $j=62,\ldots,67$ over $500$ collisions. $n$ is the collision number. Compare this figure with Fig. \ref{fig:ND}a, noting the different vertical scales.} 	\label{fig:NDNM}
\end{figure}

The normed difference for the negative zero modes is in excellent agreement with Eq. (\ref{eqn:NDNM}), as expected, and the best fit to the numerical data for the zero modes is obtained at $\beta \approx 1/192.7$. The negative non-zero modes' normed differences experience a faster growth rate than that of the zero modes. This is likely also due to numerical instability, but the negative mode normed differences nonetheless resemble the form predicted by Eq. (\ref{eqn:NDNM}). 

\subsection{Inner Products}
\label{sec:NRIP}

The form of the normalized tangent dynamics [see Eqs. (\ref{eqn:PFNTD}) and (\ref{eqn:NTDNM})] means that the inner product of the symplectic pair $\deta_{j}(t)$ and $\deta_{200-j+1}(t) = (J_{200}\deta_{j})(t)$ should grow as \begin{equation}
	\deta_j(t)\cdot\deta_{200-j+1}(t) = \frac{\beta t}{[(\beta t)^2 +1]^{1/2}}~. \end{equation}
In Fig. (\ref{fig:AD}) we present the inner product growth rates for the transverse and longitudinal modes. The zero modes are in excellent agreement with the predicted functional form, and fitting obtains a time scale $\beta = 1/196.3$. The growth rate for the other modes is in good agreement with the predicted behavior, although mode $96$ and $91$ appear to exhibit a decay in their inner product after $800$ collisions. 

\begin{figure}[htp]
	\begin{center}
	\includegraphics{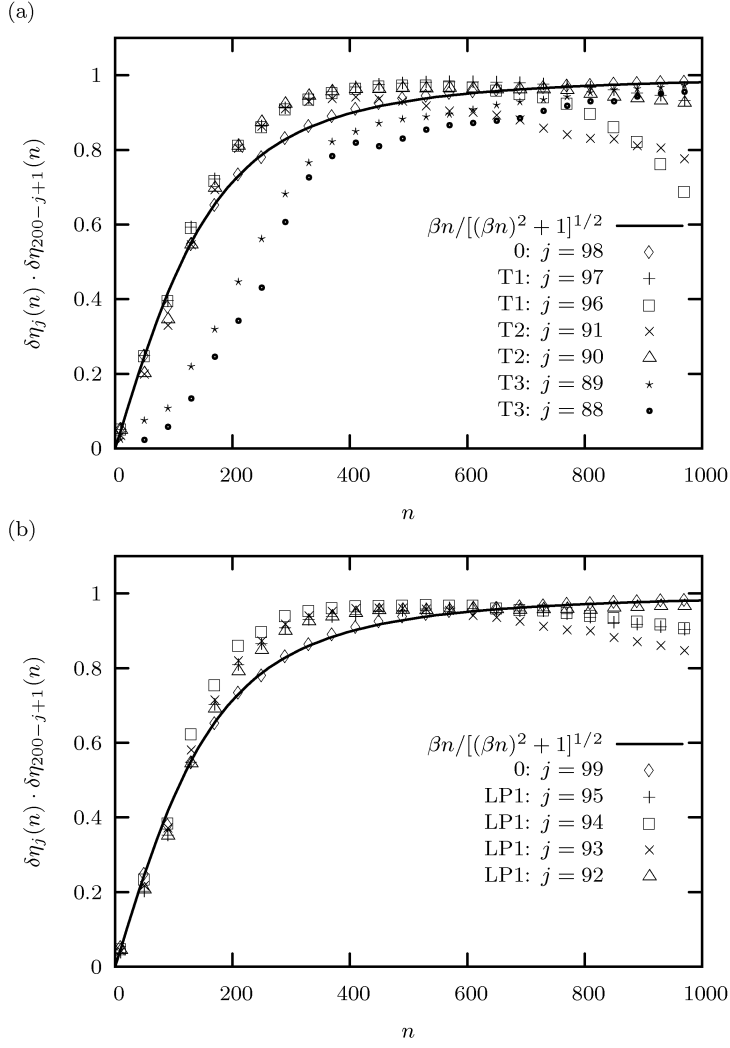}
	\end{center}
	\caption{Inner product $\deta_j(n)\cdot\deta_{200 - j +1}(n)$ for transverse (a) and longitudinal (b) modes as function of collision number $n$. The corresponding Lyapunov step label for each symplectic pair of modes is shown in the legend. For comparison the predicted growth rate $\beta n/[(\beta n)^2 +1]^{1/2}$ is also shown, fitted to data for the zero mode inner products, e.g. $\deta_{98}(n)\cdot\deta_{103}(n)$ or $\deta_{99}(n)\cdot\deta_{102}(n)$, which all coincide. The best fit for both plots is obtained at $\beta = 1/196.3$ with reduced $\chi^2 = 4.8\times10^{-5}$.} 	\label{fig:AD}
\end{figure}

The orthogonality of $\A(t)$ in Eq. (\ref{eqn:PFNTD}) also implies that positive modes should stay orthonormal to positive modes and negative modes to negative modes. We test this orthogonality explicitly by computing $\deta_i(t)\cdot\deta_j(t)$, for $88\le i\le j \le 113$. Results are shown in Table \ref{tab:MIP}. 

\begin{table}[t]
\begin{center}
	\begin{tabular}{|c|*{9}r|}
	\hline
	& \multicolumn{1}{c}{89} & \multicolumn{1}{c}{90} & \multicolumn{1}{c}{91} & \multicolumn{1}{c}{92} 	& \multicolumn{1}{c}{93} & \multicolumn{1}{c}{94} & \multicolumn{1}{c}{95} & \multicolumn{1}{c}{96} 	& \multicolumn{1}{c|}{97}\\
	\hline
	88 & -0.10 & 0.06 & -0.36 & -0.48 & -0.01 & 0.07 & 0.46 & -0.05 & -0.12 \\ 	89 & 1.00 & 0.32 & 0.20 & 0.25 &-0.17 & -0.00 & 0.25 & -0.04 & 0.01\\ 	90 & $\cdot$ & 1.00 & 0.04 & -0.10 & -0.14 & -0.33 & 0.29 & 0.01 & 0.09\\ 	91 & $\cdot$ & $\cdot$ & 1.00 & 0.16 & -0.00 & 0.12 & -0.39 & -0.09 & 0.11\\ 	92 & $\cdot$ & $\cdot$ & $\cdot$ & 1.00 & -0.16 & 0.03 & -0.06 & 0.03 & -0.03\\ 	93 & $\cdot$ & $\cdot$ & $\cdot$ & $\cdot$ & 1.00 & 0.20 & -0.19 & -0.05 & 0.04\\ 	94 & $\cdot$ & $\cdot$ & $\cdot$ & $\cdot$ & $\cdot$ & 1.00 & -0.20 & -0.04 & -0.10\\ 	95 & $\cdot$ & $\cdot$ & $\cdot$ & $\cdot$ & $\cdot$ & $\cdot$ & 1.00 & 0.04 & -0.14 \\ 	96 & $\cdot$ & $\cdot$ & $\cdot$ & $\cdot$ & $\cdot$ & $\cdot$ & $\cdot$ & 1.00 & 0.02\\ 	97 & $\cdot$ & $\cdot$ & $\cdot$ & $\cdot$ & $\cdot$ & $\cdot$ & $\cdot$ & $\cdot$ & 1.00\\ 	\hline
	\end{tabular}
	\end{center}
	\begin{center}
	\begin{tabular}{|c|*{9}r|}
	\hline
	& \multicolumn{1}{c}{105} & \multicolumn{1}{c}{106} & \multicolumn{1}{c}{107} & \multicolumn{1}{c}{108} 	& \multicolumn{1}{c}{109} & \multicolumn{1}{c}{110} & \multicolumn{1}{c}{111} & \multicolumn{1}{c}{112} 	& \multicolumn{1}{c|}{113}\\
	\hline
	104 & 0.02 & -0.06 & -0.02 & 0.01 & 0.07 & 0.01 & -0.01 & 0.06 & -0.12 \\ 	105 & 1.00 & 0.11 & -0.03 & -0.02 & 0.01 & -0.06 & 0.05 & 0.13 & 0.07 \\ 	106 & $\cdot$ & 1.00 & -0.18 & -0.23 & -0.13 & -0.07 & 0.17 & -0.26 & 0.08 \\ 	107 & $\cdot$ & $\cdot$ & 1.00 & 0.08 & -0.09 & -0.01 & -0.10 & -0.08 & -0.17\\ 	108 & $\cdot$ & $\cdot$ & $\cdot$ & 1.00 & 0.09 & -0.07 & 0.05 & 0.03 & -0.23\\ 	109 & $\cdot$ & $\cdot$ & $\cdot$ & $\cdot$ & 1.00 & 0.17 & 0.01 & 0.01 & -0.10\\ 	110 & $\cdot$ & $\cdot$ & $\cdot$ & $\cdot$ & $\cdot$ & 1.00 & 0.06 & -0.18 & 0.31\\ 	111 & $\cdot$ & $\cdot$ & $\cdot$ & $\cdot$ & $\cdot$ & $\cdot$ & 1.00 & -0.18 & 0.07 \\ 	112 & $\cdot$ & $\cdot$ & $\cdot$ & $\cdot$ & $\cdot$ & $\cdot$ & $\cdot$ & 1.00 & 0.05\\ 	113 & $\cdot$ & $\cdot$ & $\cdot$ & $\cdot$ & $\cdot$ & $\cdot$ & $\cdot$ & $\cdot$ & 1.00\\ 	\hline
	\end{tabular}
	\end{center}
	\caption{Mode inner products $\deta_i(n)\cdot\deta_j(n)$ after $n=500$ collisions, where $i$ is the row index and $j$ the column index. The column labeled (top) $88$ or (bottom) $104$ contains only the single entry $1.00$ and is therefore not shown. Top: positive mode inner products for $88 \le i\le j \le 97$. Bottom: negative mode inner products for $104 \le i\le j \le 113$.} 	\label{tab:MIP}
\end{table}

The off-diagonal inner products in Table \ref{tab:MIP} are mostly non-zero. In order to gain a sense of scale, it is instructive to compare the size of these inner products with those of the non-modes, by computing $\dxi_i(t)\cdot\dxi_j(t)$ for $61\le i \le j \le 69$. Results are shown in Table \ref{tab:NMIP}. The inner products of the non-modes appear larger than that of the modes: the average absolute value of the off-diagonal inner product for the positive modes is $0.14$ and for the negative modes is $0.09$; however for the non-modes the average is $0.31$ and standard deviation $0.18$. Note that e.g. modes 95 and 92 both seem to be rotating toward the more unstable mode 88, whilst mode 94 seems to rotate toward mode 90. Rotation toward more unstable modes is characteristic of the numerical instability in the system. Since the off-diagonal inner products of the modes are small in the scale of the non-mode inner products, it seems orthogonality of the positive (negative) modes with respect to positive (negative) modes may be preserved by the tangent dynamics up to numerical instability. 

\begin{table}[ht]
\begin{center}
	\begin{tabular}{|c|*{9}r|}
	\hline
	& \multicolumn{1}{c}{62} & \multicolumn{1}{c}{63} & \multicolumn{1}{c}{64} & \multicolumn{1}{c}{65} 	& \multicolumn{1}{c}{66} & \multicolumn{1}{c}{67} & \multicolumn{1}{c}{68} & \multicolumn{1}{c}{69} 	& \multicolumn{1}{c|}{70}\\
	\hline
	61 & 0.23 & -0.25 & -0.13 & -0.09 & -0.31 & 0.36 & -0.26 & -0.36 & 0.21 \\ 	62 & 1.00 & -0.04 & -0.10 & -0.35 & -0.64 & -0.13 & -0.15 & 0.36 &-0.08 \\ 	63 & $\cdot$ & 1.00 & 0.20 & -0.52 & -0.39 & 0.19 & -0.30 & 0.52 & 0.14 \\ 	64 & $\cdot$ & $\cdot$ & 1.00 & -0.34 & -0.18 & 0.24 & 0.48 & 0.03 & 0.39\\ 	65 & $\cdot$ & $\cdot$ & $\cdot$ & 1.00 & 0.80 & -0.29 & 0.37 & -0.46 & -0.62\\ 	66 & $\cdot$ & $\cdot$ & $\cdot$ & $\cdot$ & 1.00 & -0.34 & 0.27 & -0.29 & -0.48\\ 	67 & $\cdot$ & $\cdot$ & $\cdot$ & $\cdot$ & $\cdot$ & 1.00 & -0.00 & -0.40 & 0.71\\ 	68 & $\cdot$ & $\cdot$ & $\cdot$ & $\cdot$ & $\cdot$ & $\cdot$ & 1.00 & -0.45 & -0.07 \\ 	69 & $\cdot$ & $\cdot$ & $\cdot$ & $\cdot$ & $\cdot$ & $\cdot$ & $\cdot$ & 1.00 & -0.25\\ 	70 & $\cdot$ & $\cdot$ & $\cdot$ & $\cdot$ & $\cdot$ & $\cdot$ & $\cdot$ & $\cdot$ & 1.00\\ 	\hline
	\end{tabular}
	\end{center}
	\caption{Non-mode inner products $\dxi_i(n)\cdot\dxi_j(n)$ after $n=500$ collisions, where $i$ is the row index, $j$ the column index and $61 \le i \le j \le 69$. The column labeled $61$ contains only the single entry $1.00$ and is not therefore shown.} 	\label{tab:NMIP}
\end{table}

\subsection{Symplecticity Property}

It is also of interest to test the symplecticity property (\ref{eqn:MGLM}) by computing the sum $\log(\|\deta_j(t)\|) + \log(\|\deta_{200-j+1}(t)\|)$. Figure \ref{fig:SM} shows this sum for the modes compared to the predicted curve $1/2\log[(\beta t)^2 +1]$. Once more the zero mode data are in excellent agreement with the predicted functional form (\ref{eqn:MGLM}), as expected, while the non-zero modes appear to diverge from it. Moreover, for the longitudinal modes $92$ to $95$ the sum appears to grow linearly, rather than logarithmically, which was a key assumption in Sec. (\ref{sec:LMDLMS}). However, just as in Fig. \ref{fig:ND}, such divergence may also be merely due to numerical instability. However, one cannot completely exclude the possibility that the actual mode tangent dynamics differs from that presented in Eq. (\ref{eqn:PFNTD}). 

\begin{figure}[htp]
	\begin{center}
	\includegraphics{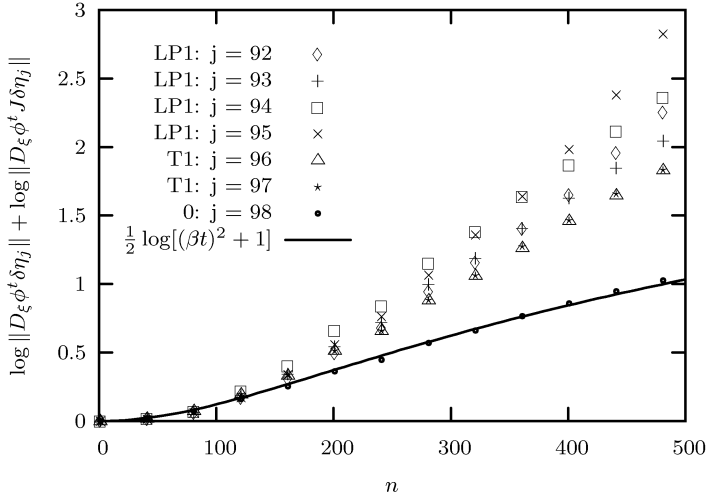}
	\end{center}
	\caption{Plots of $\log(\|\deta_j(n)\|) + \log(\|\deta_{200-j+1}(n)\|)$ as function of collision number $n$ for $j=92,\ldots,98$. The corresponding Lyapunov step label for each mode symplectic pair is shown in the legend. The zero mode symplectic pairs coincide, so that only one ($j=98$) is shown. Fitting the predicted functional form (solid line) to the zero mode data reveals excellent agreement: the best fit occurs at $\beta = 1/190.4$ with reduced $\chi^2 =3.1 \times 10^{-4}$.} 	\label{fig:SM}
\end{figure}

\section{Conclusion}
We have shown that the numerical dynamics of the Lyapunov modes in the quasi-one-dimensional system may be represented by time-modal linear combinations of an initial set of Lyapunov modes. Further, we suggested a form for the Lyapunov mode tangent dynamics, based on the form of the zero mode tangent dynamics. The inner products and normed differences of the Lyapunov modes predicted by this tangent dynamics resemble the numerical data, but the instability of the numerical method frustrates a more meaningful analysis. Our suggested tangent dynamics implies that the Lyapunov modes have well-defined Lyapunov exponents in the positive time limit. 

\begin{acknowledgements}
The authors are grateful to C Angstmann and T Chung for discussions. \end{acknowledgements}

\appendix

\section{Theoretical Summary}
\label{sec:TF}
\subsection{Tangent Space Dynamics}
\label{sec:TFTSD}
Suppose we have an $r$-dimensional system of $N$ particles, with phase flow $\flow{t}:M \to M \simeq \mathbb{R}^{2rN}$. Classification of the stability of a tangent vector $\dxi \in T_{\xi}M$ is achieved by following theorem originally due to Oseledec \cite{Ose68,ER85}. \begin{theorem} (Multiplicative Ergodic Theorem) Let $\flow{t}:M\to M$ be a (not necessarily invertible) differentiable phase flow, which preserves the Liouville measure. Then for almost all $\xi \in M$ (with respect to this measure) the stability matrix \begin{equation}
\label{eqn:FSM}
\Lambda_{\xi,+} \equiv \lim_{t \to \infty}\Big[\big(\tanflow{\xi}{t}\big)^T\tanflow{\xi}{t}\Big]^{1/2t} \end{equation}
exists and there exists a positive integer $l$ and distinct real numbers $\lam{j}$ such that $\Lambda_{\xi,+}$ has eigenvalues $\exp(\lam{j})$ for $j= 1,\ldots,l$, where $\lam{1} > \ldots > \lam{l}$. Let $\Uf{j}{\xi}$ be the eigenspace of $\Lambda_{\xi,+}$ corresponding to eigenvalue $\exp(\lam{j})$, and let \begin{equation}
	\label{eqn:DPVS}
	\Vf{j}{\xi} \equiv \Uf{j}{\xi}\oplus\cdots\oplus\Uf{l}{\xi}~, \end{equation}
$\Vf{l+1} \equiv \{\bm{0}\}$. Then furthermore, \begin{equation}
\label{eqn:FTAB}
\Vf{j}{\xi} - \Vf{j+1}{\xi} = \bigg\{\dxi \in T_{\xi}M : \lim_{t \to \infty}\frac{1}{t}\log\frac{\| \tanflow{\xi}{t}\dxi \|}{\| \dxi \|} = \lam{j}\bigg\}~. \end{equation}
\end{theorem}
As previously, the numbers $\lam{j}$ are called Lyapunov exponents, and the ordered set $\{\lam{j}\}$ is called the Lyapunov spectrum. We have included the denominator in the logarithm of Eq. (\ref{eqn:FTAB}) to indicate explicitly that the Lyapunov exponent describes a ratio of tangent vector lengths. This denominator contributes a zero term to the exponent in the limit $t \to \infty$, so we neglect it henceforth. 

The stability matrix $\Lambda_{\xi,+}$ is clearly symmetric, so the eigenspaces $\Uf{j}{\xi}$ form an orthogonal decomposition of the tangent space. That is, $T_{\xi}M = \Uf{1}{\xi}\oplus\cdots\oplus\Uf{l}{\xi}$ and $\langle\dxi,\deta\rangle = 0$ for $\dxi \in \Uf{j}{\xi}$, $\deta \in \Uf{k}{\xi}$ and $j \not=k$, where we use the usual inner product here - the dot product - which defines the norm $\|\cdot\|$. It follows from this decomposition and the definition of the spaces $\Vf{j}{\xi}$ that $T_{\xi}M = \Vf{1}{\xi} \supset \ldots \supset \Vf{l}{\xi} \supset \Vf{l+1}{\xi} = \{\bm{0}\}$. 

According to the theorem, the sets $\Vf{j}{\xi} - \Vf{j+1}{\xi}$ partition $T_{\xi}M$ according to the Lyapunov exponents: each non-zero tangent vector has an unique associated Lyapunov exponent. This means that we may write \begin{align}
\Vf{j}{\xi}
& = \bigcup_{k=j}^l\Big(\Vf{k}{\xi} - \Vf{k+1}{\xi}\Big) \cup\{\bm{0}\}\notag \\ & = \bigg\{\dxi \in T_{\xi}M : \lim_{t \to \infty}\frac{1}{t}\log\| \tanflow{\xi}{t}\dxi \| \le \lam{j}\bigg\}\cup\{\bm{0}\}~, \label{eqn:FVP} \end{align}
and call $\Vf{j}{\xi}$ the $(l-j+1)$-th most stable subspace of $T_{\xi}M$ in the positive time limit, or the positive $(l-j+1)$-th stable subspace for short. From the definition of the spaces $\Vf{j}{\xi}$ in the theorem it also follows immediately that \begin{equation}
\label{eqn:FEALV}
\Uf{j}{\xi} - \{\bm{0}\} \subset \Vf{j}{\xi} - \Vf{j+1}{\xi}~, \end{equation}
and hence the non-zero elements of the eigenspaces have associated Lyapunov exponent $\lam{j}$, as would be expected. Further, orthogonality of the eigenspaces and Eq. (\ref{eqn:DPVS}) implies \begin{equation}
\label{eqn:FELVS}
\Uf{j}{\xi} = \Vf{j}{\xi}\cap\big(\Vf{j+1}{\xi}\big)^{\perp}~, \end{equation}
so that the eigenspace $\Uf{j}{\xi}$ is that subspace orthogonal to the positive $(l-j)$-th stable subspace but contained within the positive $(l-j+1)$-th stable subspace. 

\subsection{Hamiltonian Properties}
\label{sec:TFHP}
The phase flow of the particle system is Hamiltonian, and consequently the composition of phase flows exhibits the (homomorphism) property \begin{equation}
\label{eqn:PFH}
\flow{t}\circ\flow{s} = \flow{t + s}~.
\end{equation}
The phase flow is clearly invertible, with inverse $(\flow{t})^{-1} = \flow{-t}$ and $\flow{0}$ is the identity map. (As such, the set $\{\flow{t}\}_{t \in \mathbb{R}}$ is an Abelian group that is isomorphic to the real numbers.) We have then from Eq. (\ref{eqn:PFH}) and the chain rule that the tangent flow is a co-cycle, that is \begin{equation}
\label{eqn:TFCP}
\tanflow{\flow{s}(\xi)}{t}\tanflow{\xi}{s} = \tanflow{\xi}{t+s}~. \end{equation}
The tangent flow is therefore invertible, with inverse $(\tanflow{\xi}{t})^{-1} = \tanflow{\flow{t}(\xi)}{-t}$ and $\tanflow{\xi}{0} = I$, the identity matrix. 

A Hamiltonian phase flow preserves the Liouville measure on $M$ \cite{Arn89}, so that the phase flow under consideration satifies the hypotheses of the Multiplicative Ergodic Theorem. Invertibility of the phase flow means that we may apply the theorem also to the inverse $\flow{-t}$, which is equivalent to contemplating a switch in the direction of time. It follows from the theorem that the stability matrix \begin{equation}
\label{eqn:RSM}
\Lambda_{\xi,-} \equiv \lim_{t \to -\infty}\Big[\big(\tanflow{\xi}{t}\big)^T\tanflow{\xi}{t}\Big]^{1/2|t|} \end{equation}
also exists, and $\Lambda_{\xi,-}$ must have the inverse eigenvalues to $\Lambda_{\xi,+}$, so that it has eigenvalues $\exp(-\lam{j})$, for $j=1,\ldots,l$. As we expect, the Lyapunov exponents of $\flow{-t}$ are then the opposite sign to those of $\flow{t}$ but have the same multiplicity, denoted $\mbox{mult}(j)$. We let $\Ur{j}{\xi}$ be the eigenspace corresponding to eigenvalue $\exp(-\lam{j})$, and then $T_{\xi}M = \Ur{1}{\xi}\oplus\cdots\oplus\Ur{l}{\xi}$ is another orthogonal decomposition of the tangent space. In general, note that $\Uf{j}{\xi} \not= \Ur{j}{\xi}$, but we do have that $\dim \Uf{j}{\xi} = \mbox{mult}(j) = \dim \Ur{j}{\xi}$. 

The $j$-th most stable subspace in the negative time limit $\Vr{j}{\xi}$, called the negative $j$-th stable space, is now well-defined by the theorem. Since $-\lam{1} < \ldots < -\lam{l}$ are the Lyapunov exponents, we must have\footnote{Note that since the order of the Lyapunov exponents has been reversed, we have reversed the indexing of these spaces so that the $(l-j+1)$-th stable space is $\Vr{l-j+1}{\xi}$ rather than $\Vr{j}{\xi}$.} \begin{equation}
\label{eqn:DRSS}
\Vr{j}{\xi} \equiv \Ur{1}{\xi}\oplus\cdots\oplus\Ur{j}{\xi}~, \end{equation}
and then the tangent space is partitioned into the sets \begin{equation}
	\label{eqn:RPTS}
\Vr{j}{\xi} - \Vr{j-1}{\xi} = \bigg\{\dxi \in T_{\xi}M : \lim_{t \to -\infty}\frac{1}{|t|}\log\| \tanflow{\xi}{t}\dxi \| = -\lam{j}\bigg\}~, \end{equation}
where $T_{\xi}M = \Vr{l}{\xi} \supset\ldots\supset\Vr{1}{\xi}\supset\Vr{0}{\xi} \equiv \{\bm{0}\}$. From Eq. (\ref{eqn:DRSS}) we have \begin{equation}
\label{eqn:RELVS}
\Ur{j}{\xi} = \Vr{j}{\xi} \cap \big(\Vr{j-1}{\xi}\big)^{\perp}~, \end{equation}
similarly to Eq. (\ref{eqn:FELVS}). The non-zero elements of the eigenspaces $\Ur{j}{\xi}$ has associated exponent $-\lam{j}$ analogously to Eq. (\ref{eqn:FEALV}), and note also that, just as in Eq. (\ref{eqn:FVP}), here \begin{equation}
\label{eqn:RVP}
\Vr{j}{\xi} = \bigg\{\dxi \in T_{\xi}M : \lim_{t \to -\infty}\frac{1}{|t|}\log\| \tanflow{\xi}{t}\dxi \| \le -\lam{j}\bigg\}\cup\{\bm{0}\}~.
\end{equation}

We call the elements of the eigenspaces $\Uf{j}{\xi}$ ($\Ur{j}{\xi}$) the Lyapunov vectors, each of which has an associated Lyapunov exponent in the positive (negative) time limit. We may clearly choose an orthonormal basis of $T_{\xi}M$ such that each element of the basis is an element of an eigenspace $\Uf{j}{\xi}$ ($\Ur{j}{\xi}$), just as claimed in Section \ref{sec:LMLEV}. Further, if any of the Lyapunov exponents have multiplicity greater than unity, then we have many choices of such a basis and only the eigenspace itself is well-defined. 

\subsection{Covariant Subspaces}
\label{sec:TFCS}
Suppose the tangent vector $\dxi \in \Vf{j}{\xi} - \Vf{j+1}{\xi}$, so that it has corresponding Lyapunov exponent $\lam{j}$. If $\eta = \flow{s}(\xi)$ then $\deta = \tanflow{\xi}{s}\dxi \in T_{\eta}M$ should also have Lyapunov exponent $\lam{j}$, since the Lyapunov exponents are independent of time. This property is guaranteed by the co-cycle property of the tangent flow and its consequent invertibility. It can be shown that \begin{equation}
\label{eqn:ISD}
\Vf{j}{\flow{s}(\xi)} = \tanflow{\xi}{s}\Vf{j}{\xi}~, \end{equation}
for $s$ finite. Further, since the $\Vf{j}{\xi}$ are closed subspaces, the equality also holds in either limit $s \to \pm\infty$. A subspace satisfying Eq. (\ref{eqn:ISD}) for all real $s$ and $s \to \pm\infty$ is called a covariant subspace, and an identical result holds for the $\Vr{j}{\xi}$. Note that the set $\Vf{j}{\xi} - \Vf{j+1}{\xi}$ is not covariant since the equality does not necessarily hold in the limit $s \to -\infty$, as this set is not closed. 

For a non-zero $\dxi \in \Vf{j}{\xi}$, we have from Eq. (\ref{eqn:ISD}) that $\tanflow{\xi}{-s}\dxi \in \Vf{j}{\flow{-s}(\xi)}$. Substituting this into the inequality in Eq. (\ref{eqn:FVP}) and applying the successive transformations $t \to t-s$ then $s \to -(s+t)$, which together are the map $t \to -s$, one finds that \begin{equation}
\label{eqn:FVP2}
\Vf{j}{\xi} \subseteq \bigg\{\dxi \in T_{\xi}M : \lim_{t \to -\infty}\frac{1}{|t|}\log\| \tanflow{\xi}{t}\dxi \| \ge -\lam{j}\bigg\}\cup\{\bm{0}\}~, \end{equation}
and similarly
\begin{equation}
\label{eqn:RVP2}
\Vr{j}{\xi} \subseteq \bigg\{\dxi \in T_{\xi}M : \lim_{t \to +\infty}\frac{1}{|t|}\log\| \tanflow{\xi}{t}\dxi \| \ge \lam{j}\bigg\}\cup\{\bm{0}\}~. \end{equation}

Equation (\ref{eqn:FEALV}) implies the eigenspace $\Uf{j}{\xi}$ ($\Ur{j}{\xi}$) has associated Lyapunov exponent $\lam{j}$ ($-\lam{j}$) in the positive (negative) time limit. However, Eqs. (\ref{eqn:FVP2}) and (\ref{eqn:RVP2}) unfortunately imply that in the negative (positive) time limit we can only say that a non-zero element of $\Uf{j}{\xi}$ ($\Ur{j}{\xi}$) has exponent $\lambda \ge -\lam{j}$ ($\lambda \ge \lam{j}$). Nevertheless, comparison of Eqs. (\ref{eqn:FVP}), ({\ref{eqn:RVP}), (\ref{eqn:FVP2}) and (\ref{eqn:RVP2}) suggests that the intersection of the spaces $\Vf{j}{\xi}$ and $\Vr{j}{\xi}$ produces a subspace with well-defined Lyapunov exponent in both the positive or negative time limit. That is, let \begin{equation}
\W{j}{\xi} \equiv \Vf{j}{\xi}\cap\Vr{j}{\xi}~. \end{equation}
Then
\begin{equation}
\W{j}{\xi} = \bigg\{\dxi \in T_{\xi}M: \lim_{t \to \pm\infty}\frac{1}{|t|}\log\| \tanflow{\xi}{t}\dxi \| = \pm\lam{j}\bigg\}\cup\{\bm{0}\}~, \end{equation}
and it can be proven (see e.g. Ref. \cite{Rue79}) that \begin{equation}
\label{eqn:OS}
\Vr{j}{\xi} = \W{1}{\xi}\oplus\cdots\oplus\W{j}{\xi}~, \end{equation}
so that $T_{\xi}M = \W{1}{\xi}\oplus\cdots\oplus\W{l}{\xi}$. This is called the Oseledec decomposition of the tangent space, and $\W{j}{\xi}$ is called an Oseledec space. Note that the Oseledec spaces are not orthogonal like the $\Ur{j}{\xi}$, but are clearly covariant spaces unlike the $\Ur{j}{\xi}$. From Eqs. (\ref{eqn:OS}) and (\ref{eqn:RELVS}) it follows that $\dim \Ur{j}{\xi} = \dim\W{j}{\xi}$. 

\subsection{Numerical Scheme}
\label{sec:TFNS}

The Lyapunov exponents are calculated via the numerical scheme of Benettin \cite{BGGS80} and Shimada \cite{SN79}. A byproduct of this scheme is a orthonormal basis of Lyapunov vectors that are elements of the eigenspaces $\Ur{j}{\xi}$. The numerical scheme depends upon the following well-known result, which we state as a proposition. We call the map $\tanflow{\xi}{t}\dxi/\|\tanflow{\xi}{t}\dxi\|$ the normalized tangent flow. 

\emph{Proposition} A randomly chosen non-zero, tangent vector converges for $t \to +\infty$ under the normalized tangent flow to the most stable space in the negative time limit, $\Vr{1}{\xi}$. That is for non-zero $\dxi \in T_{\xi}M$ and $t \to \infty$, \begin{equation}
\frac{\tanflow{\xi}{t}\dxi}{\|\tanflow{\xi}{t}\dxi\|} \to \mbox{proj}\bigg(\Vr{1}{\flow{t}(\xi)}, \frac{\tanflow{\xi}{t}\dxi}{\|\tanflow{\xi}{t}\dxi\|}\bigg)~, \end{equation}
in the norm $\|\cdot\|$, where $\mbox{proj}(V,x)$ denotes the projection of the vector $x$ onto the subspace $V$. 

The proposition can be proved by use of the Oseledec decomposition of the tangent space, the invariance property of the Oseledec spaces and the fact that these spaces possess well-defined, ordered Lyapunov exponents (see e.g. \cite{EP98}). The idea here is that a randomly chosen tangent vector converges to its projection onto the space $\Vr{1}{} = \W{1}{}$ under the normalized tangent flow, but it does not converge at all in the norm $\|\cdot\|$ under the usual tangent flow. 

Let
\begin{equation}
	M(j) = \bigg\{k\in \mathbb{Z}: \sum_{i=1}^{j-1}\mbox{mult}(i) < k \le \sum_{i=1}^{j}\mbox{mult}(i)\bigg\}~. \end{equation}
The proposition may be generalized to show that for $k$ randomly chosen, nonzero, linearly independent tangent vectors $\dxi_i$, \begin{equation}
\label{eqn:MTVC}
\lim_{t \to \infty}\Vr{j-1}{\flow{t}(\xi)} \subset \lim_{t \to \infty} \mbox{span}\bigg\{\frac{\tanflow{\xi}{t}\dxi_i}{\|\tanflow{\xi}{t}\dxi_i\|}\bigg\}_{i=1}^k \subseteq \lim_{t \to \infty}\Vr{j}{\flow{t}(\xi)}~, \end{equation}
where $j$ is such that $k \in M(j)$.

We now fix $t$ and define $\{\deta^n_i\}$ recursively as the Gram-Schmidt orthonormalized set generated by $\{[\tanflow{\flow{[n-1]t}(\xi)}{t}]\deta^{n-1}_i\}$, where $\{\deta^0_i\}$ is the orthonormalized set of the tangent vector set $\{\dxi_i\}$. Note that the tangent flow is not orthogonal. It follows then from Eqs. (\ref{eqn:DRSS}) and (\ref{eqn:MTVC}) that \begin{equation}
\lim_{n \to \infty} \deta^n_k \in \lim_{n\to \infty}\Ur{j}{\flow{nt}(\xi)} \end{equation}
for $k \in N(j)$. In other words, the numerical scheme operates as follows: if we randomly choose a set of $2rN$ linearly independent tangent vectors, then apply the tangent flow to these vectors whilst regularly Gram-Schmidt orthonormalizing them, we produce in the positive time limit an orthonormal basis of Lyapunov vectors. These Lyapunov vectors have well-defined Lyapunov exponents only in the negative time limit. 

\subsection{Further Hamiltonian Properties} \label{sec:TFFHP}
A Hamiltonian phase flow also preserves the symplectic structure of the phase space $M$, so that the tangent flow is a symplectic matrix. Hence \begin{equation}
\label{eqn:TFSM}
(\tanflow{\xi}{t})^TJ_{2rN}\tanflow{\xi}{t} = J_{2rN}~, \end{equation}
where $J_{2rN}$ is defined in Eq. (\ref{eqn:DSMJ}), and we put $J \equiv J_{2rN}$ henceforth. It is well-known that, due to this symplecticity, there is always a zero Lyapunov exponent and all other exponents are paired signwise with equal multiplicity such that \begin{equation}
\label{eqn:HRLE}
\lam{l-j+1} = -\lam{j}~.
\end{equation}
Hence the number of distinct exponents, $l$, is always odd and $\lam{[l+1]/2} = 0$. Henceforth we let $m \equiv (l+1)/2$, and we call the associated eigenspaces $\Ur{m}{\xi}$ or $\Uf{m}{\xi}$ the zero eigenspaces. 

The stability matrices $\Lambda_{\xi,\pm}$ (\ref{eqn:FSM},\ref{eqn:RSM}) are thus also symplectic. It can then be shown from Eq. (\ref{eqn:HRLE}) and the definition of the $U^{(j)}_{\xi,\pm}$ as eigenspaces of $\Lambda_{\xi,\pm}$, that \begin{equation}
\label{eqn:HRES}
U^{(j)}_{\xi,\pm} = JU^{(l-j+1)}_{\xi,\pm} \end{equation}
and hence for the zero eigenspaces
\begin{equation}
\label{eqn:IJZMS}
U^{(m)}_{\xi,\pm} = JU^{(m)}_{\xi,\pm}~. \end{equation}
Note that for all $\dxi \in T_{\xi}M$, the dot product $\dxi\cdot J\dxi = 0$. We deduce $\dim U^{(m)}_{\xi,\pm} = \mbox{mult}(m)$ must be even, since if $\dxi$ is an element of an orthogonal basis for $U^{(j)}_{\xi,\pm}$, then by Eq. (\ref{eqn:IJZMS}) so is $J\dxi$. 

An important consequence of Eqs. (\ref{eqn:HRLE}) and (\ref{eqn:HRES}) is that the Lyapunov spectrum and an orthonormal basis of Lyapunov vectors is fully described by considering only positive Lyapunov exponents and the Lyapunov vectors corresponding to these. That is, for a orthonormal set of Lyapunov vectors $\{\dxi_k\}$, $k = 1,\ldots,2rN$, we may choose $\dxi_{2rN - j + 1} = J\dxi_j$ for $j =1,\ldots,rN$. 

\subsection{Symmetries and Zero Modes}
\label{sec:TFSZM}
If we presume that the dynamics of the system under consideration has full (translational) spatial and time symmetry, then there are consequently $r+1$ tangent vectors $\dxi_i$, $i = 1,\ldots,r+1$ such that \begin{equation}
	\label{eqn:DSZM}
	\tanflow{\xi}{t}\dxi_i = \dxi_i~.
\end{equation}
The first $r$ tangent vectors of these correspond to a uniform translation of the $N$ particles in the $i$-th spatial direction. With reference to Eq. (\ref{eqn:DTVC}) these are $\dxi_i = 1/N^{1/2}\big(0,e_i,\ldots,e_i\big)$, where $e_i \in \mathbb{R}^r$ is the $i$-th unit basis vector repeated $N$ times and $0$ is the $rN$ dimensional zero vector. Symmetry in time means that the system also possesses time translational invariance, and hence the velocity unit tangent vector at state $\xi = (p,q)$, $\dxi_{r+1} = 1/\|p\|\big(0,p\big)$, satisfies $\tanflow{\xi}{t}\dxi_{r+1} = \dxi_{r+1}$. 

By Noether's theorem \cite{Arn89}, the spatial and time symmetries of the Hamiltonian dynamics give rise to conserved quantities, namely momentum and energy respectively. Hence we have a further $r$ tangent vectors corresponding to a momentum shift in the $i$-th direction $\dxi_{i+r+1} = 1/N^{1/2}\big(e_i,\ldots,e_i,0\big)$, for $i = 1,\ldots,r$. These tangent vectors have dynamics \begin{equation}
	\label{eqn:DPZM}
	\tanflow{\xi}{t}\dxi_{i+r+1} = \dxi_{i+r+1} + \beta t\dxi_i~, \end{equation}
where $\beta>0$ is some constant, such that they are generalized eigenvectors of $\tanflow{\xi}{t}$ with algebraic multiplicity $2$. Further, a shift in energy effectively raises or lowers the temperature of the system, and hence we have a tangent vector $\dxi_{2r+2} = 1/\|p\|\big(p,0\big)$ which has dynamics \begin{equation}
	\label{eqn:DEZM}
	\tanflow{\xi}{t}\dxi_{2r+2} = \dxi_{2r+2} + \beta t\dxi_{r+1}~, \end{equation}
corresponding to linear separation in time. Note that we have written all the $\dxi_i$ with unit norm. 

Now, $\dxi_{i+r+1} = -J\dxi_i$ for all $i = 1,\ldots,r+1$. Applying this relation together with the symplecticity of the tangent flow to Eqs. (\ref{eqn:DSZM}), (\ref{eqn:DPZM}) and (\ref{eqn:DEZM}) one finds that for $i =1,\ldots,r+1$ \begin{equation}
	\label{eqn:ADSZM}
		\big(\tanflow{\xi}{t})^T\dxi_i = \dxi_i + J\beta t\dxi_i~, \end{equation}
It follows by the definition of the stability matrices in the Multiplicative Ergodic Theorem with Eqs. (\ref{eqn:DSZM}) and (\ref{eqn:ADSZM}) that \begin{align}
\Lambda_{\xi,\pm}\dxi_{i}
& = \lim_{t \to \pm\infty}\Big[\big(\tanflow{\xi}{t}\big)^T\tanflow{\xi}{t}\Big]^{1/2|t|}\dxi_{i}\notag\\ & = \lim_{t \to \pm \infty}\big(I + \beta tJ\big)^{1/2|t|}\dxi_{i}\notag\\ & = \dxi_{i}~,
\end{align}
for $i = 1,\ldots,r+1$, and a similar result holds for $i = r+2,\ldots,2r+2$. By definition of $\Uf{m}{\xi}$ and $\Ur{m}{\xi}$ as a zero eigenspace of $\Lambda_{\xi,+}$ and $\Lambda_{\xi,-}$ respectively, we have \begin{equation}
\label{eqn:SZMCE}
\mbox{span} \Big\{\dxi_i\Big\}_{i=1}^{2r+2} \subseteq \Ur{m}{\xi}\cap\Uf{m}{\xi}~. \end{equation}
We therefore call $\dxi_i$ the zero modes. 

\subsection{Equivalence of Zero Spaces}
\label{sec:TFEZS}

Empirical results confirm that the linear combinations of the $\dxi_i$ in App. \ref{sec:TFSZM} are indeed the zero modes generated by the numerical scheme of App. \ref{sec:TFNS}. Moreover, one finds the multiplicity of $\lam{m}$, $\mbox{mult(m)} = 2s+2$, for $s$ the number of unbroken spatial symmetries in the system of interest. It follows that $\dim\Uf{m}{\xi} = \dim\Ur{m}{\xi} = \dim(\mbox{span}\{\dxi_i\})$. Since the zero modes are linearly independent, then with reference to Eq. (\ref{eqn:SZMCE}) it must be that \begin{equation}
\label{eqn:CZESZM}
\Ur{m}{\xi} = \mbox{span}\Big\{\dxi_i\Big\}_{i=1}^{2r+2} = \Uf{m}{\xi}~, \end{equation}
In other words, the zero eigenspaces coincide and their non-zero elements consequently have a well-defined Lyapunov exponent $\lambda = 0$ in either the positive or negative time limit. 

>From the definition of the subspace $\Vf{j}{\xi}$ in the Multiplicative Ergodic Theorem, Eq. (\ref{eqn:CZESZM}) implies that $\Ur{m}{\xi} \subset \Vf{m}{\xi}$. Hence $\Ur{m}{\xi} \subseteq \Vf{m}{\xi}\cap\Vr{m}{\xi} \equiv \W{m}{\xi}$. However, $\dim \W{m}{\xi} = \dim\Ur{m}{\xi}$ so it must be that 
\begin{equation}
	\label{eqn:EZS}
\W{m}{\xi} = \Ur{m}{\xi} = \Uf{m}{\xi}~, \end{equation}
or in other words the zero eigenspaces and the zero Oseledec subspace coincide. This result is a direct consequence of the structure of the zero modes and the empirically verified multiplicity of the zero exponent.

\section{Symplecticity of Numerical Dynamics} \label{sec:ASND}
Consider the numerical dynamics $\Omega^t = \Nd(\Omega)$ defined in Sec. \ref{sec:LMPND}, and let $\Omega = (\ldots|\delta\omega_j|\ldots)$ be an orthogonal $2n\times 2n$ matrix of Lyapunov vectors such that $\delta\omega_{2n-j+1} = J \delta \omega_j$. Note that this choice of $\delta \omega_j$ coincides with a possible choice of an ordered set of Lyapunov vectors generated by the numerical scheme (see App. \ref{sec:TFFHP}). We now show that $\Nd$ is a symplectic operator, in the sense that \begin{equation}
	\label{eqn:ASP}
	J\Nd(\delta \omega_j) = \Nd(J\delta\omega_j)~, \end{equation}
for $j=1,\ldots,2n$ and $J \equiv J_{2n}$ as defined in Eq. (\ref{eqn:DSMJ}). 

Firstly, we may consider the numerical dynamics to consist of evolution under the tangent flow $\tanflow{\xi}{t}$ followed by the Gram-Schmidt orthonormalization procedure(see App \ref{sec:TFNS}). Letting $\Omega(t) = \tanflow{\xi}{t}\Omega$, then the numerical dynamics may be written as $\Omega^t = \Omega(t)R^{-1}(t)$ where $R(t)$ is the upper triangular matrix of the so-called $QR$ factorization of $\Omega(t)$. Hence we have \begin{equation}
	\Nd(\Omega) = \tanflow{\xi}{t}\Omega R^{-1}(t)~, \end{equation}
as a decomposition of the numerical dynamics. $\Nd$ is explicitly dependent on the choice of $\Omega$ and is implicitly dependent on the ordering of the $\delta \omega_j$ due to the iterative nature of the Gram-Schmidt procedure, so it is not a linear map, although the matrix $\Nd(\Omega)$ is clearly orthogonal. 

Now, $(\delta\omega_j)^t = \Nd(\delta \omega_j) = \tanflow{\xi}{t}\Omega R^{-1}(t)\Omega^T\delta\omega_j$. Defining the matrix \begin{equation}
	\label{eqn:ADND}
	N(t) \equiv \tanflow{\xi}{t}\Omega R^{-1}(t)\Omega^T~, \end{equation}
we also have that $\Nd(J\delta \omega_j) = \Nd(\delta \omega_{2n-j+1}) = N(t)J\delta\omega_j$. Since the $\delta \omega_j$ form a basis of the tangent space, Eq. (\ref{eqn:ASP}) is then equivalent to $JN(t) = N(t)J$. Since $N(t)$ is orthogonal, it therefore suffices to show that $N(t)$ is symplectic in order to show Eq. (\ref{eqn:ASP}). 

Let $\mathcal{R}(t) = \Omega R(t)\Omega^T$. Then the orthogonality of $N(t)$, i.e. $N^{-1}(t) = N^T(t)$, together with Eq. (\ref{eqn:ADND}) implies that $\mathcal{R}(t)^T\mathcal{R}(t) = (\tanflow{\xi}{t})^T\tanflow{\xi}{t}$. It follows that $\mathcal{R}(t)^T\mathcal{R}(t)$ is symplectic due to the symplecticity of the tangent flow. That is, \begin{equation}
	\label{eqn:ASCR}
	\mathcal{R}(t)^T\mathcal{R}(t)J\mathcal{R}(t)^T\mathcal{R}(t) = J~. \end{equation}
By hypothesis, we may write
\begin{equation}
	\Omega = \Big(\Delta\Big|J\Delta\tilde{I}\Big) \end{equation}
where $\Delta$ is an $2n\times n$ matrix with orthonormal columns that satisfies $\Delta^TJ\Delta = 0$, and $\tilde{I}$ is the $n\times n$ antidiagonal matrix with $1$s on the antidiagonal and zeros elsewhere. It then follows from this choice that \begin{equation}
	\label{eqn:AMSO}
	\Omega^TJ\Omega = \tilde{J}~,\ \mbox{where}\ \tilde{J} = \begin{pmatrix} 0&-\tilde{I} \\ \tilde{I}& 0 \end{pmatrix}~. \end{equation}
Hence Eq. (\ref{eqn:ASCR}) implies
\begin{equation}
	\label{eqn:AMSRTR}
	R^T(t)R(t)\tilde{J}R^T(t)R(t) = \tilde{J}~. \end{equation}

By Eq. (\ref{eqn:ADND}), in order for $N(t)$ to be symplectic we require $\mathcal{R}^TJ\mathcal{R} = J$. From Eq. (\ref{eqn:AMSO}) this requirement is equivalent to \begin{equation}
	\label{eqn:AMSR}
	R(t)^T\tilde{J} R(t) = \tilde{J}.
\end{equation}
It is therefore sufficient to show Eq. (\ref{eqn:AMSR}) in order to show that $N(t)$ is symplectic and hence Eq. (\ref{eqn:ASP}) holds. 

One can show via some algebra that Eq. (\ref{eqn:AMSRTR}) together with the upper triangular nature of $R(t)$ produces Eq. (\ref{eqn:AMSR}). In detail, we write \begin{equation}
	R(t) = \begin{pmatrix} A & B \\ 0 & C \end{pmatrix} \end{equation}
where $A$, $C$ are themselves $n\times n$ upper-triangular matrices with strictly positive values on the diagonal. Eq. (\ref{eqn:AMSRTR}) provides restrictions on $A$, $B$ and $C$ so that one finds \begin{equation}
	R(t)\tilde{J}R^T(t) = \begin{pmatrix} 0 & -A\tilde{I}C^T\\ C\tilde{I}A^T & 0 \end{pmatrix}~. \end{equation}
$A\tilde{I}C^T$ is an upper antitriangular matrix (an upper triangular matrix multiplied by $\tilde{I}$) with strictly positive values on the antidiagonal,and one finds further that $A\tilde{I}C^T$ is also orthogonal. The only possibility is $A\tilde{I}C^T = \tilde{I}$, so since $\tilde{J}^2 = -I$, Eq. (\ref{eqn:AMSR}) holds and so does Eq. (\ref{eqn:ASP}). 

\bibliography{modes}

\end{document}